\documentclass[prb,twocolumn,superscriptaddress]{revtex4-2}

\usepackage{amsmath,amssymb}
\usepackage[pdftex]{graphicx}
\usepackage{bm}
\usepackage{bbm}
\usepackage{hyperref}
\usepackage{color} 
\usepackage{srcltx} 
\usepackage{newtxtext,newtxmath}

\allowdisplaybreaks

\begin{document}

\title{Edge-induced pairing states in a Josephson junction through a spin-polarized quantum anomalous Hall insulator
}
\date{\today}

\author{Ryota Nakai}
\affiliation{Institute for Materials Research,
Tohoku University, Sendai 980-8577, Japan}
\author{Kentaro Nomura}
\affiliation{Institute for Materials Research,
Tohoku University, Sendai 980-8577, Japan}
\affiliation{Center for Spintronics Research Network, 
Tohoku University, Sendai 980-8577, Japan}
\author{Yukio Tanaka}
\affiliation{Department of Applied Physics, 
Nagoya University, Nagoya 464-8603, Japan}

\begin{abstract}
Despite the robustness of the chiral edge modes of quantum Hall systems against the superconducting proximity effect, Cooper pairs can penetrate into the chiral edge channels and carry the Josephson current in an appropriate setup.
In our work, the Josephson junction of a spin-polarized quantum anomalous Hall insulator (QAHI) with a Chern number $\nu=1$ connecting conventional superconductors is studied from the perspective of pairing symmetry consistent with the chiral edge mode.
Induced pairing states are equal-spin triplet, a combination of the even- and odd-frequency components, nonlocally extended, and have a finite momentum $2k_F$.
The signature of the equal-spin triplet pairings is confirmed via the dependence on the interface-magnetization direction, and that of the finite-momentum pairing states via the spatial profile of the anomalous Green's function.
In the presence of disorder, the robustness of the chiral edge mode leads to high sensitivity of the critical current and the equilibrium phase difference to disorder configurations, which is resulting from the interference of current-carrying channels.
The numerical calculations on a lattice model are also examined by a simplified analytical model.
\end{abstract}

\maketitle


\section{Introduction}

Quantum Hall phases \cite{prange87} and conventional superconductor phases are electronic states of matter characterized by different orders, topological and spontaneously-symmetry-broken orders, while both are immune to disorder and support non-dissipative electric current.
The quantized Hall current in a quantum Hall insulator is carried by conducting edge channels, the number of which coincides with the Chern number defined by the bulk electronic states.
These edge channels consist of unidirectionally flowing electronic system (chiral edge mode), which is unique to quantum Hall edges in that they cannot be realized in closed one-dimensional electronic systems due to the quantum anomaly.

Supercurrent in a superconductor is carried by pairs of electrons called the Cooper pairs.
While the superconducting order is measured by the pair potential $\Delta_{\sigma\sigma'}(t,x)$,
the amount of pairing states is measured by the pair amplitude (or the anomalous correlation) $-i\langle \mathcal{T}c_{\sigma}(t,x)c_{\sigma'}(0,0)\rangle$, where $c_{\sigma}(t,x)$ is the annihilation operator of the electron with a spin $\sigma$ at temporal and spatial coordinate $(t,x)$ and $\mathcal{T}$ is the time-ordering operator.
The pair potential and pair amplitude are categorized by the spin configuration, angular momentum, and symmetry regarding time reversal, which as a whole obey the Pauli principle, that is, the permutation of two electrons changes the sign of the pair potential (amplitude) \cite{berezinskii74,balatsky92,linder19}.
Among them, conventional superconductors indicate superconductors having an even-frequency/spin-singlet/even-parity (ESE) pair potential.
On the other hand, odd frequency indicates that the pair potential (amplitude) is odd under time reversal (and thus odd under the sign change of the frequency). 
Although realization of the odd-frequency pair potential in the bulk of superconductors is still under debate \cite{fominov15}, the odd-frequency pair amplitude appears ubiquitously at the surface or the interface of even-frequency superconductors \cite{linder19}.
Specifically, in the heterostructure of a normal metal and a conventional superconductor, ESE pairs are transformed into a combination of ESE and odd-frequency/spin-singlet/odd-parity (OSO) pairs during the tunneling into a normal metal, due to the breaking of translational symmetry \cite{tanaka07,tanaka07-2,tanaka12}.

When a ferromagnet is attached to a conventional superconductor, the penetration depth of singlet pairs is limited by a length determined by the exchange coupling \cite{bergeret01,bergeret05,eschrig07}, and, furthermore, singlet pairs are completely excluded in the limit of a half metal since pairings between opposite spins are prohibited. 
However, in the presence of a mechanism to flip the spin at the interface, singlet pairs are transformed into equal-spin triplet pairs, which penetrate even into a half metal at long range \cite{eschrig03,asano07,eschrig08,galaktionov08,beri09}.
The resulting triplet pairs are even-frequency odd-parity (ETO) pairs in addition to odd-frequency even-parity (OTE) pairs since translational symmetry is broken at the interface \cite{bergeret01,bergeret05,tanaka07-2}.

Topological materials have unique electronic states on their boundary, and thus the heterostructure with superconductors can have a different functionality from non-topological materials \cite{alicea12,beenakker13}.
Especially, the proximity-induced odd-frequency pairings in topological materials have been studied in 3d topological insulators \cite{yokoyama12,black-schaffer12,black-schaffer13,lu15,burset15,breunig18,lu18}, quantum spin Hall insulators (QSHI) \cite{crepin15,cayao17,kuzmanovski17,keidel18,fleckenstein18,cayao20}, Rashba metals \cite{reeg15,ebisu16,bobkova17,cayao18,cayao20}, and Weyl semimetals \cite{parhizgar20,dutta19,dutta20}.

The focus of this study is the Josephson junction of a conventional superconductor and a spin-polarized quantum anomalous Hall insulator (QAHI).
Experimentally, heterojunctions of quantum Hall systems and superconductors have been reported e.g. in \cite{amet16,lee17,matsuo18,seredinski19,zhao20,gul20}.
In quantum Hall systems, symmetry of induced pairing states depends on the nature of chiral edge modes.
When chiral edge modes are spin-degenerate, such as in a spin-degenerate quantum Hall state with a Chern number $\nu=2$ \cite{ma93,zyuzin94,ishikawa99,giazotto05,stone11,huang17,alavirad18,gavensky20-2}, the Josephson current is carried by singlet (ESE and OSO) pairs.
Similar pairing states occur in a single chiral edge mode of a spin-unpolarized QAHI \cite{lian16,sakurai17}, in which the spin-polarization axis depends on the direction of the boundary.  
However, the situation is different in a spin-polarized single chiral edge mode \cite{fisher94,vanostaay11,michelsen20}.
In general, a spin-polarized chiral edge mode is not a good conductor of the Josephson current due to (i) the chirality by which backward Andreev reflection does not occur, (ii) the Pauli exclusion principle by which equal spins cannot be present at the same position and time, and (iii) spin polarization which prohibits opposite-spin pairings to flow into the edge \cite{fisher94}.

In this work, we study the DC Josephson effect of a junction comprising of a spin-polarized QAHI and conventional $s$-wave superconductors by putting emphasis on symmetry of induced pairing states. 
Since the spin-polarized chiral edge mode is a less transparent conductor of conventional ESE pairs due to the above-mentioned three factors, induced pairings are shown to be (i) equal-spin triplet, (ii) a combination of even and odd frequencies, and have (iii) a finite momentum (the Fulde-Ferrell state \cite{fulde64}).
In addition, (iv) the anomalous correlation extends nonlocally throughout the QAHI edge due to the robustness of the chiral edge mode against the proximity effect and disorder.
By comparing with the Josephson effect through a QSHI, it will become clear that these pairing states are unique to the spin-polarized chiral edge mode.
Notice that in our setup the bulk of a QAHI is not influenced by the proximity effect, which may cause a topological phase transition to a topological superconductor phase \cite{qi10,he17,lian18}.
In addition, we consider only conventional superconductors, while the Josephson junction between unconventional ones has been studied in \cite{gavensky20}.

This paper is organized as follows.
A lattice model of a superconductor/QAHI/superconductor (S/QAHI/S) Josephson junction is introduced in Sec.~\ref{sec:latticemodel}, and is used to calculate the anomalous correlation and the Josephson current through a clean and dirty QAHI in Sec.~\ref{sec:numerical}.
From analytical perspective, a model of a Josephson junction through a chiral edge mode as an low-energy model of the S/QAHI/S junction is introduced in Sec.~\ref{sec:chiraledgemodel}, and the corresponding physical quantities are estimated perturbatively in Sec.~\ref{sec:analytical}.
Finally, the conclusion is given in Sec.~\ref{sec:conclusion}.

\section{A lattice model for numerical calculations}
\label{sec:latticemodel}

\begin{figure}
 \centering
 \includegraphics[width=84mm]{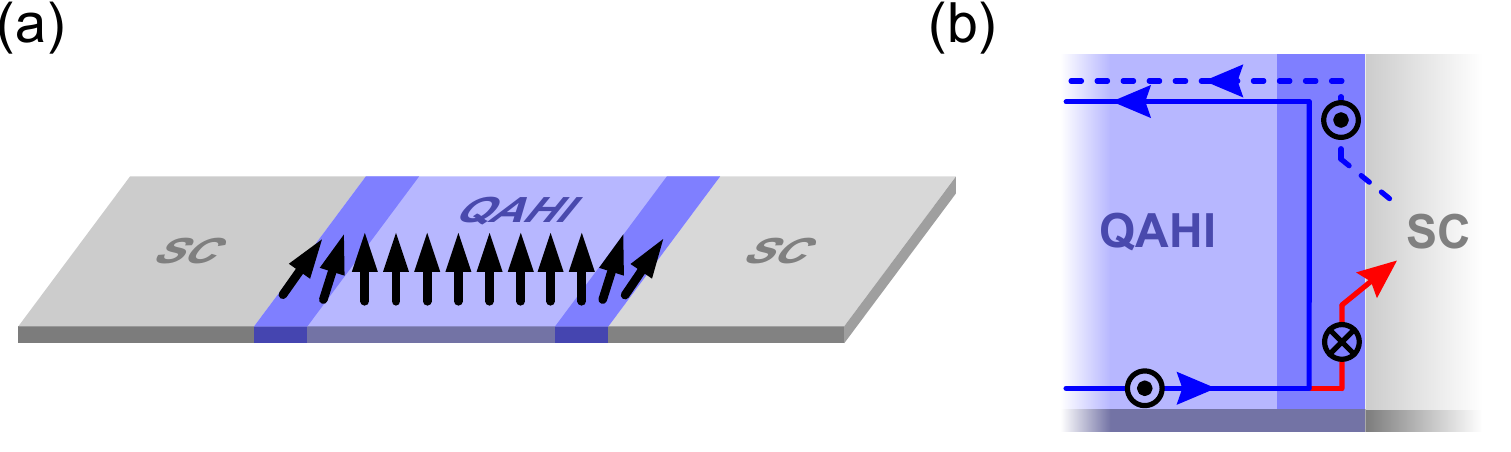}
 \caption{(a) A setup for the lattice model of a S/QAHI/S Josephson junction. Near the interfaces with superconductors, the regions of which are referred to as the left and right interface layers and are shown by darker purple in the figure, the exchange field (black arrow) is tilted from the bulk direction.
 (b) A mechanism of the equal-spin crossed Andreev reflection. The blue solid (dashed) line represents the spin-up electron (hole), and the red solid line represents the spin-down electron. The arrows indicate the direction in which electrons and holes flow.
 \label{fig:sqahis}}
\end{figure}
Consider a Josephson junction of a spin-polarized QAHI sandwiched by two $s$-wave superconductors [Fig.~\ref{fig:sqahis} (a)].
First, we model a spin-polarized QAHI with a Chern number $\nu=1$ by the Bernevig-Hughes-Zhang (BHZ) model \cite{bernevig06} of a QSHI coupled with an exchange field $\bm{M}$.
The Fourier transform of the Hamiltonian defined on the square lattice is given by $H_\text{QAHI}=H_\text{BHZ}+H_\text{exchange}$, where
\begin{align}
 &\mathcal{H}_\text{BHZ}
 =
 \frac{\epsilon_s+\epsilon_p}{2}
 -
 (t_s-t_p)
 (\cos k_x+\cos k_y) \notag\\
 &\qquad+
 \left[
 \frac{\epsilon_s-\epsilon_p}{2}
 -
 (t_s+t_p)
 (\cos k_x+\cos k_y)
 \right]s^z \notag\\
 &\qquad+
 2t_{sp}
 \left(
 s^y\sin k_x + \sigma^zs^x\sin k_y
 \right), \\
 &\mathcal{H}_\text{exchange}
 =
 -M_x\sigma^x
 -
 M_y\sigma^ys^z
 -
 M_z\sigma^zs^z.
 \label{eq:hamiltonian_exchange}
\end{align}
Here, $\sigma$ and $s$ are Pauli matrices for spin and orbital ($s$ and $p$) degrees of freedom, respectively.
Without the exchange field ($\bm{M}=0$), the BHZ model shows a trivial insulator phase when $\epsilon_s-\epsilon_p>4(t_s+t_p)$, while it shows a QSHI phase when $\epsilon_s-\epsilon_p<4(t_s+t_p)$ and $t_{sp}\neq 0$ \cite{bernevig13}.
In the presence of the exchange field, the QSHI phase, where both spin components are in quantum Hall phases with opposite Chern numbers, is turned to a QAHI phase, where one of the spin components is in trivially insulating phase, via a topological phase transition.
When we fix $t_s=t_p>0$ and $0<\epsilon_s=-\epsilon_p<4t_s$ and the exchange field is along the $z$ direction ($\bm{M}=M_z\hat{z}$), the BHZ model shows a QSHI phase when $M_z\in [\epsilon_s-4t_s,-\epsilon_s+4t_s]$, while QAHI phases when $M_z\in[-\epsilon_s-4t_s,\epsilon_s-4t_s]$ and $[-\epsilon_s+4t_s,\epsilon_s+4t_s]$, where the spin of the chiral edge mode is polarized along $-\hat{z}$ and $\hat{z}$ directions, respectively.
Notice that the exchange term (\ref{eq:hamiltonian_exchange}) is made to respect four-fold rotational symmetry of the BHZ model: the $s$ orbital has spin 1/2 and the $p$ orbital has total spin 3/2, and thus the BHZ model has symmetry of
$
 \mathcal{H}_\text{BHZ}(k_y,-k_x)
 =
 R\mathcal{H}_\text{BHZ}(k_x,k_y)R^{-1}
$,
where $R=e^{-i\pi\sigma^z/4}(1+s^z)/2+e^{-3i\pi\sigma^z/4}(1-s^z)/2$. 

A conventional $s$-wave superconductor is modeled by
\begin{align}
 \mathcal{H}_\text{SC}
 =
 \begin{pmatrix}
  \xi_k & i\sigma^y\Delta \\
  -i\sigma^y\Delta^{\ast} & -\xi_{-k}^{\ast}
 \end{pmatrix},
\end{align}
where the $2\times 2$ matrix represents the Nambu space of the electron and hole degrees of freedom and $\xi_k=[2t_\text{SC}(2-\cos k_x-\cos k_y)-\mu]\sigma^0$.

Near the interface between the QAHI and two superconductors, the exchange field $\bm{M}$ is slightly tilted from the direction of the bulk field in order to introduce triplet pairings into the spin-polarized chiral edge mode of the QAHI [Fig.~\ref{fig:sqahis} (a)].
At the interface of the QAHI and the superconductors, the number of electron bands mismatches (4 electron bands in the QAHI while 2 electron bands in the superconductors). 
Here, we assume that an electron in the superconductors tunnels to $s$- and $p$-orbitals of the QAHI by the same tunneling amplitude without flipping the spin.
The tunneling between the QAHI and the superconductors is made to respect the four-fold rotational symmetry so that the both interfaces behave in the same way.

The formation of equal-spin pairing states is resulting from the equal-spin crossed Andreev reflection through which spin-up electrons in one boundary are reflected as spin-up holes in the other boundary [Fig.~\ref{fig:sqahis} (b)]. 
This phenomenon consists of the following two processes \cite{braude07,eschrig07,asano07-2,eschrig08,beri09}. First, the tilt of the spin-polarization axis in the interface layer mixes spin-up and spin-down electrons, and then the usual Andreev reflection transforms spin-down electrons into spin-up holes [Fig.~\ref{fig:sqahis} (b)].
Equivalently, in the reversed order, the Andreev reflection transforms incoming spin-up electrons into spin-down holes, from which the interface layer generates spin-up holes.

\section{Numerical results}
\label{sec:numerical}

In this section, numerical calculations performed with the lattice model are presented.
The parameters are fixed as $\epsilon_s-\epsilon_p=6$, $t_s=t_p=1$, $t_{sp}=0.5$, $t_{\text{SC}}=1$, $\mu=2$, $|\Delta_\text{R}|=|\Delta_\text{L}|=0.5$, and the tunneling amplitude $t_{\text{tunnel}}=0.5$.
The chemical potential is defined by $\mu=\epsilon_s-3$, where $\mu=0$ corresponds to the center of the bottom of the conduction bands and the top of the valence bands.
Notice that we employ relatively large pair potential to make the observation easier.
The exchange field in the bulk of the QAHI is $\bm{M}=(0,0,2)$ along the $z$ direction, while that in the QSHI is $\bm{M}=(0,0,0)$.
Near the interface with superconductors, the first and second layers have the direction of the exchange field tilted from the $z$ direction by polar angles of $0.1\pi$ and $0.05\pi$, respectively, and the polarization axis of the right interface lies within the $M_x$-$M_z$ plane [Fig.~\ref{fig:sqahis}, see also Fig.~\ref{fig:phaseazimuthalangle} (a)].  
The width of the junction is fixed to be 20 and the length of two superconductors to be 20, which is much longer than the coherence length $v/|\Delta|\sim 2$.
The length of the QAHI is $L=30$ except for the study of the length dependence in Sec.~\ref{sec:numerical_dcjosephsoneffect}.

\subsection{Induced pairings}
\label{sec:pairing}

\begin{figure}
 \centering
 \includegraphics[width=80mm]{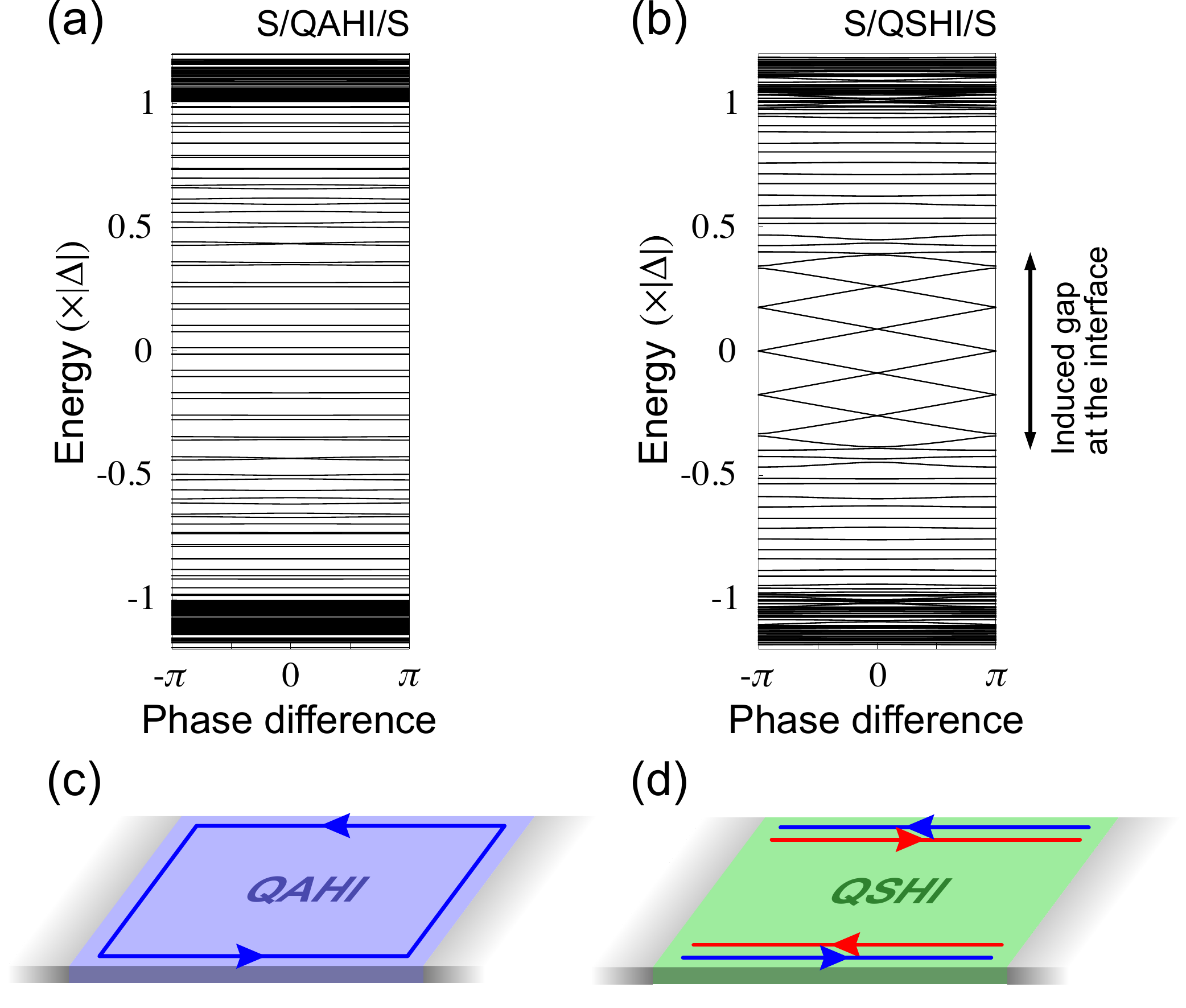}
 \caption{The energy spectrum of the whole Josephson junction is shown as a function of the phase difference between two superconductors for (a) a S/QAHI/S junction and (b) a S/QSHI/S junction.
The corresponding geometry and the edge modes are shown for (c) a S/QAHI/S junction and (d) a S/QSHI/S junction, where the blue (red) lines represent spin-up (-down) modes and the arrows represent the direction in which both electrons and holes flow.
 \label{fig:spectralflow}}
\end{figure}

First, we consider how the chiral edge mode is affected by the superconducting proximity effect.
The chiral edge mode is known to be robust against the proximity effect in the sense that the linear dispersion cannot be gapped, which can be seen in the spectrum of the chiral edge mode.
Fig.~\ref{fig:spectralflow} (a) represents the change of the energy spectrum of a S/QAHI/S Josephson junction as a function of the phase difference between the left and right superconductors.
The energy spectrum within the pair potential $|\Delta_{\text{R(L)}}|$ is equally spaced, and looks almost insensitive to the phase difference.
Since the chiral edge mode extends along the whole boundary of the QAHI [Fig.~\ref{fig:spectralflow} (c)], the inter-level spacing is determined by the perimeter of the QAHI.
This result is in stark contrast to the energy spectrum of the Josephson junction via a QSHI (S/QSHI/S) [Fig.~\ref{fig:spectralflow} (b)], in which the proximity effect induces a gap of $\sim 0.4|\Delta|$, and the energy spectrum of the bound states is largely affected by the change of the phase difference. 
The ingap modes within the induced gap are bound to the top and bottom boundaries where the proximity effect is absent [Fig.~\ref{fig:spectralflow} (d)], and thus the interlevel-spacing is determined by the length of the QSHI (or equivalently, the distance between two superconductors).

From the perspective of the Andreev reflection, the chirality of a QAHI edge prohibits a hole to be reflected locally, that is, a spin-up electron flowing along the bottom boundary is reflected as a spin-up hole on the top boundary [Fig.~\ref{fig:spectralflow} (c)], which is known as the crossed Andreev reflection.
On the other hand, since a QSHI has spin-up and spin-down edge modes flowing in the opposite direction, the Andreev reflection occurs in the ordinary way, that is, separately in the bottom and top boundaries [Fig.~\ref{fig:spectralflow} (d)], which is consistent with the gapped interface by the induced pair potential.

\begin{figure}
 \centering
 \includegraphics[width=84mm]{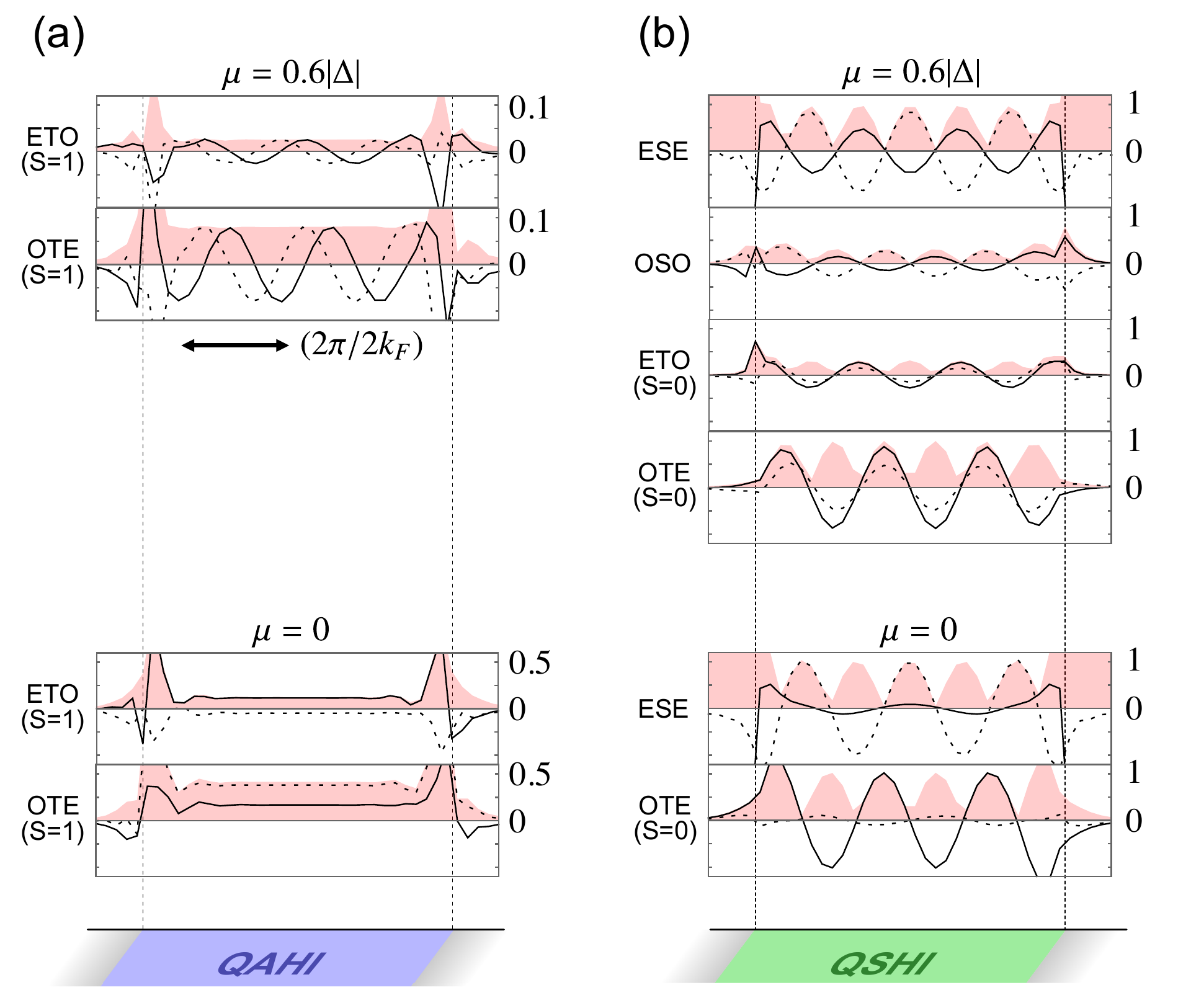}
 \caption{(a) Spatial profile of the anomalous Green's function ($F^\text{ESE/OTE}_{\bm{r}\bm{r}}$ and $F^\text{OSO/ETO}_{\bm{r}\bm{r}+\hat{e}_x}$) along (a) the S/QAHI/S junction boundary and (b) the S/QSHI/S junction boundary is shown for $\mu/|\Delta|=0,0.6$ (bottom to top) and $\omega/|\Delta|=0.6$. 
 In each figure, the real (imaginary) part of the anomalous Green's function is shown by the solid (dotted) black line, and the absolute value by the red shaded region. 
 The length of the solid arrow below the $\mu=0.6|\Delta|$ data in (a) represents the period of the anomalous Green's function.
 \label{fig:agf}}
\end{figure}

When the equal-spin crossed Andreev reflection occurs in a QAHI, triplet pairs are induced along the boundary.
The amount of the pairing states is measured by the anomalous part of the Green's function calculated by the inverse of $\omega-\mathcal{H}\pm i\delta$, where $\delta=|\Delta_\text{R/L}|/100$ is used in the following.
The anomalous Green's function is exponentially small inside the gapped bulk of the QAHI, and the only non-vanishing components in the QAHI are those between an spin-up electron and hole along the boundary.
In the following, only the $s$-orbital part is focused since two orbitals are qualitatively equivalent.
The equal-spin triplet anomalous Green's function is decomposed into symmetric and anti-symmetric parts with respect to the sign change of the frequency $\omega\to-\omega$ and the permutation of the positions as \cite{linder19}
\begin{align}
 F^\text{ETO/OTE}_{\bm{r}\bm{r}'}(\omega)
 =&
 [F_{\bm{r}\bm{r}'}^\text{R}(\omega)]_{\uparrow s,\uparrow s}
 \mp
 [F_{\bm{r}'\bm{r}}^\text{R}(\omega)]_{\uparrow s,\uparrow s} \notag\\
 &\pm
 [F_{\bm{r}\bm{r}'}^\text{A}(-\omega)]_{\uparrow s,\uparrow s}
 -
 [F_{\bm{r}'\bm{r}}^\text{A}(-\omega)]_{\uparrow s,\uparrow s},
 \label{eq:etootedefinition}
\end{align}
where $F^\text{R/A}_{\bm{r}\bm{r}'}$ is the retarded (advanced) anomalous Green's function between two points $\bm{r}$ and $\bm{r}'$.
The real-, imaginary-part, and the absolute value of even-parity components at the same site $F^\text{ESE/OTE}_{\bm{r}\bm{r}}$ and odd-parity components between neighboring sites $F^\text{OSO/ETO}_{\bm{r}\bm{r}+\hat{e}_x}$ along the junction boundary are shown for the chemical potential $\mu/|\Delta|=0$ and $0.6$, and the frequency $\omega/|\Delta|=0.6$ in Fig.~\ref{fig:agf} (a). 
As can be seen from the figure, the absolute value is almost constant while the real- and imaginary part of the anomalous Green's function have spatial oscillation.
The periodicity of the oscillation coincides with $2\pi/2k_F(=\pi v/\mu)$, where $v$ is the Fermi velocity of the chiral edge mode and in our model the same as the hopping amplitude $t_s=t_p=1$.
This is an evidence that the Cooper pair has momentum $2k_F$, that is, two electrons with both the Fermi momentum $k_F$ are paired, which has been pointed out in \cite{ishikawa99} for singlet pairings in a $\nu=2$ spin-degenerate quantum Hall system.
The finite momentum pairing is common to pairings in S/ferromagnet/S junctions \cite{sellier03,oboznov06}, while the alternation of 0 and $\pi$ junctions by the length of the weak link cannot be seen in our case.
The presence of both even- and odd-frequency components is consistent with the breaking of translational symmetry by the finite-momentum pairing, which transforms the ETO into OTE component and vice versa.

For the case of the S/QSHI/S junction, the anomalous Green's function along the boundary is also shown in Fig.~\ref{fig:agf} (b). There are four non-vanishing components away from the interface, that is, ESE and OSO components, and opposite-spin ETO and OTE components. 
The absolute value shows a stationary wave between two superconductors, and its period is approximately inversely proportional to the frequency $\omega$, but not dependent on the Fermi momentum as in the case of the S/QAHI/S junction.
This result is consistent with the fact that electrons in the QSHI boundary make pairs in the conventional way, that is, a spin-up electron with the momentum $k_F$ makes a pair with a spin-down electron with the momentum $-k_F$.

\subsection{Nonlocal correlation}
\label{sec:nonlocalcorrelation}
\begin{figure}
 \centering
 \includegraphics[width=84mm]{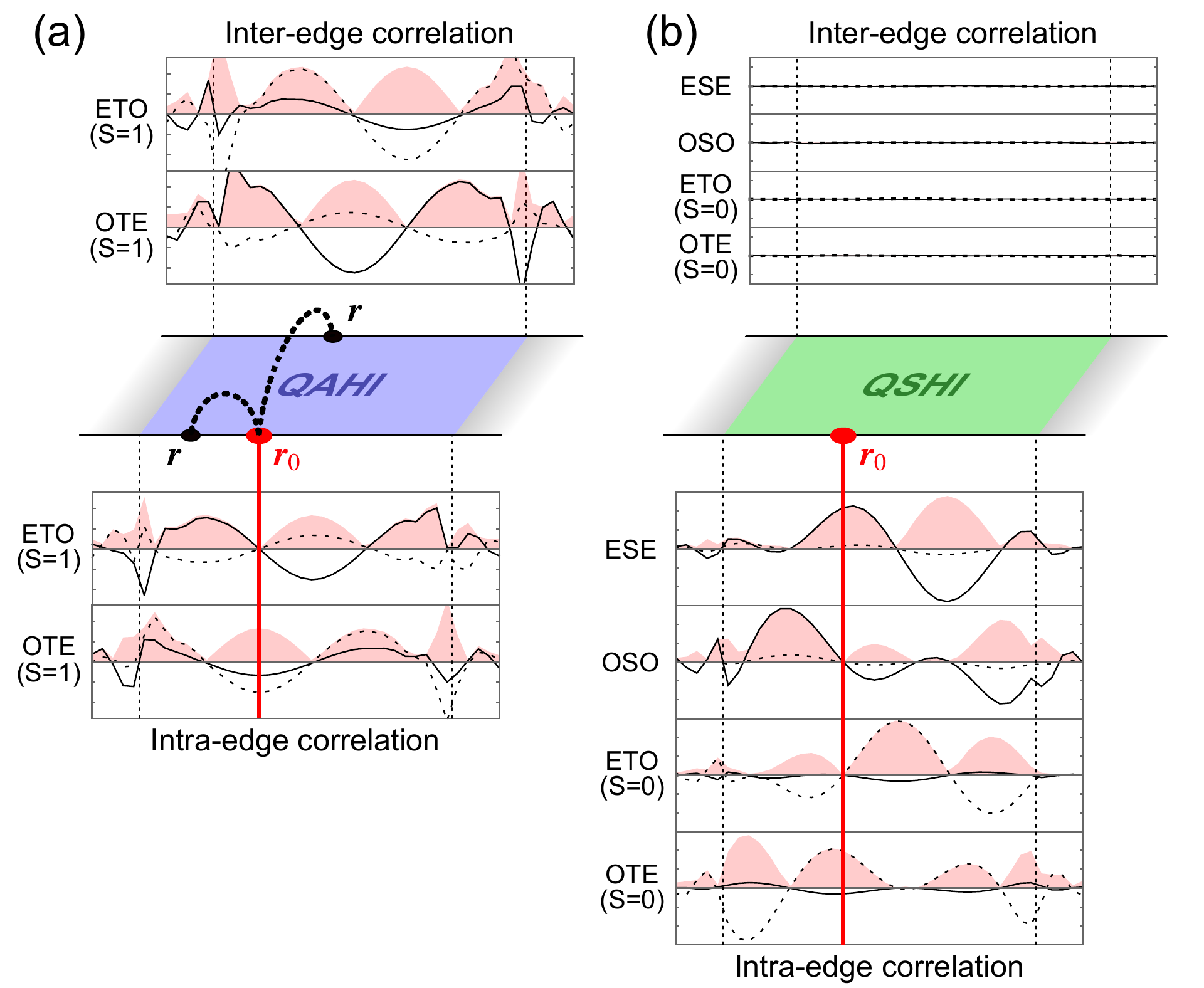}
 \caption{Nonlocal anomalous Green's function $F_{\bm{r}\bm{r_0}}$ from a fixed point $\bm{r}_0$ (shown by red circles) is shown as a function of $\bm{r}$ along the top and bottom boundaries, respectively, in (a) a S/QAHI/S junction and (b) a S/QSHI/S junction in the same manner as Fig.~\ref{fig:agf}.
 \label{fig:nonlocalagf}}
\end{figure}

Since the crossed Andreev reflection at the QAHI/S interface occurs between the top and bottom boundaries [Fig.~\ref{fig:sqahis} (b)], electrons and holes have a correlation across the boundaries in a nonlocal way.
The nonlocal anomalous Green's function in a S/QAHI/S junction between a fixed point $\bm{r}_0$ on the bottom boundary and a point $\bm{r}$ along the top and bottom boundaries is shown in Fig.~\ref{fig:nonlocalagf} (a) for $\mu/|\Delta|=0$ and $\omega/|\Delta|=0.6$.
The anomalous Green's function is finite even at the opposite boundary ($\bm{r}$ on the top boundary).
When $\mu=0$, the anomalous Green's function $F_{\bm{r}\bm{r}_0}$ is a periodic function of $\bm{r}$, whose period is inversely proportional to the frequency $\omega$.
We have also examined that the nonlocal correlations persist in the presence of disorder due to the robustness of the chiral edge mode (not shown in this paper).

Fig.~\ref{fig:nonlocalagf} (b) is the nonlocal anomalous Green's function in a S/QSHI/S junction for $\omega/|\Delta|=0.2$, which is below the induced gap ($\sim 0.4|\Delta|$) at the QSHI/S interface, and $\mu/|\Delta|=0.6$.
The anomalous Green's function extends within the same boundary since the top and bottom boundaries are separated by the gapped interface regions.
This result is consistent with an intuitive picture given in Fig.~\ref{fig:spectralflow} (c) and (d).

\subsection{DC Josephson effect}
\label{sec:numerical_dcjosephsoneffect}

Next, we consider the DC Josephson effect induced by the phase difference $\varphi$ between the pair potential of two superconductors.
When the absolute value of the pair potential is the same ($|\Delta_\text{R}|=|\Delta_\text{L}|$), the Josephson current is estimated from the free energy $F$ of the junction by $I=(2e/\hbar)dF/d\varphi$ \cite{beenakker91}, and in the limit of vanishing temperature $T\to 0$, the expression is simplified as $I=(2e/\hbar)dE/d\varphi$, where $E=\sum_{\epsilon_n<0}\epsilon_n$ is the ground-state energy. 

\begin{figure}
 \centering
 \includegraphics[width=84mm]{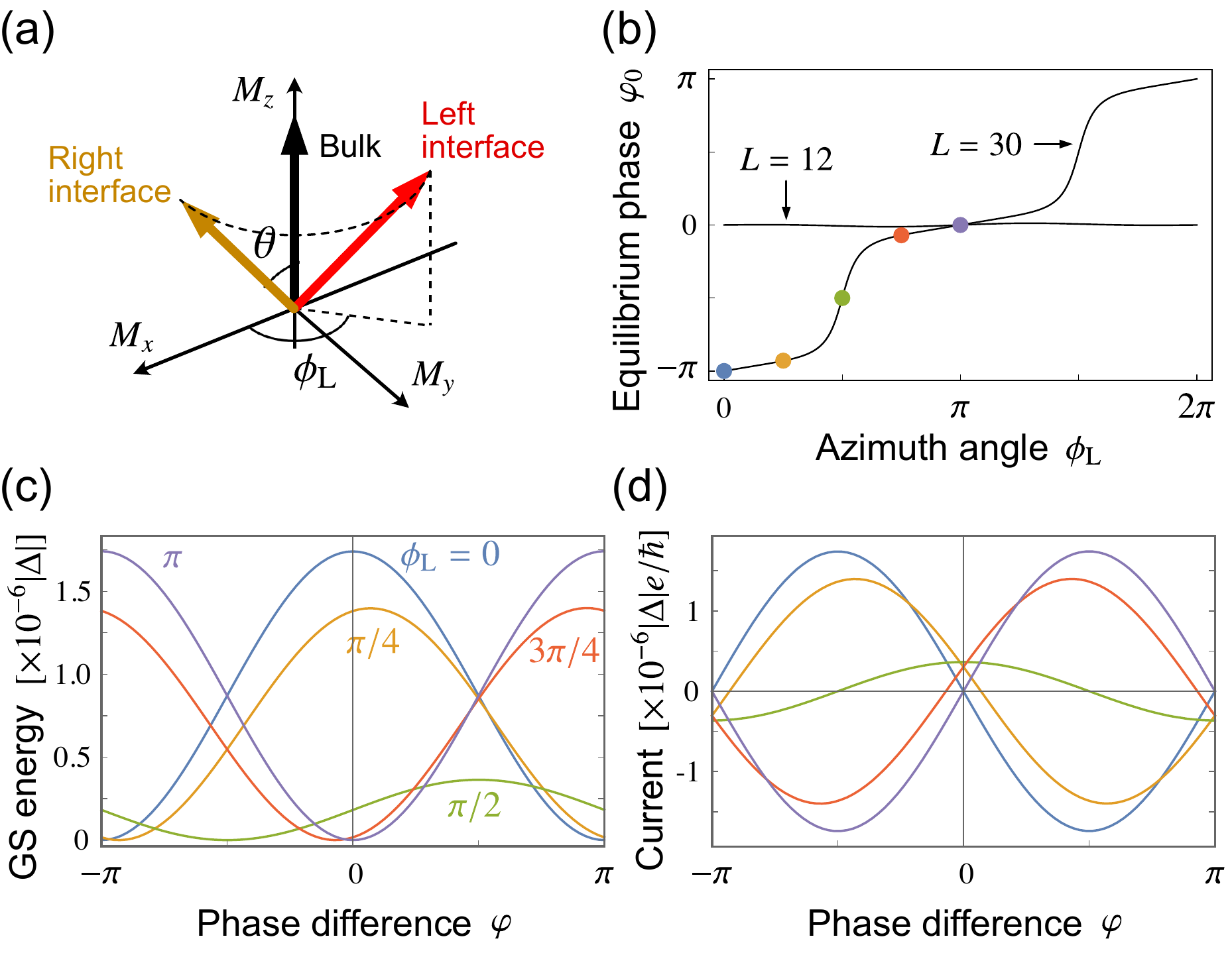}
 \caption{(a) The direction of the exchange field in the bulk (black), the right interface (yellow), and the left interface with the azimuth angle $\phi_\text{L}=3\pi/4$ (red). (b) Azimuth-angle $\phi_L$ dependence of the equilibrium phase difference of a S/QAHI/S junction is shown for a QAHI length $L=12,30$.  (c) The ground state energy and (d) the current-phase relation as a function of the phase difference are shown for the azimuth angle $\phi_\text{L}=0,\pi/4,\pi/2,3\pi/4,\pi$ and $L=30$.
Colors of the lines in (c) and (d) correspond to the dots with the same colors in (b). 
 \label{fig:phaseazimuthalangle}}
\end{figure}

When the Josephson current is carried by the equal-spin triplet pairs, the equilibrium phase difference where the ground state energy is minimized is continuously dependent on the angle formed by the direction of the exchange field of the bulk, left, and right interfaces [Fig.\ref{fig:phaseazimuthalangle} (a)] \cite{braude07,eschrig07,asano07-2,eschrig08,beri09}.
The equilibrium phase difference is either 0 or $\pi$ when the three directions are coplanar, and it is continuously changed by $2\pi$ when the direction of the exchange field of one of the interface is rotated around the bulk one.
The azimuth-angle dependence of the equilibrium phase difference in junctions with a QAHI length $L=12$ and $30$ is shown in Figs.~\ref{fig:phaseazimuthalangle} (b). 
For $L=30$, the ground state energy and the Josephson current as a function of the phase difference $\varphi$ are shown in Figs.~\ref{fig:phaseazimuthalangle} (c) and (d), respectively, for the azimuth angle of the left interface $\phi_L=0, \pi/4, \pi/2, 3\pi/4,$ and $\pi$.
The current-phase relation is almost always sinusoidal: $I=I_c\sin(\varphi-\varphi_0)$, where $I_c$ is the critical current and $\varphi_0$ is the equilibrium phase difference \cite{buzdin05,bergeret05}. 
Exceptional cases occur when accidental near-zero-energy modes cross $\epsilon=0$ during the change of the phase difference from $\varphi=0$ to $2\pi$. 
However, for a long junction these exceptions can be neglected since the variation of each eigenmode near $\epsilon=0$ is far smaller than the inter-level spacing [see Fig.~\ref{fig:spectralflow} (a)].
Thus the following calculations are performed under the assumption of the sinusoidal current-phase relation to reduce the computational cost.
When $L=30$, the equilibrium phase $\varphi_0$ varies continuously from $-\pi$ to $\pi$ during the change of the azimuth angle $\phi_\text{L}$ of the left interface from $0$ to $2\pi$.
On the other hand, for $L=12$, the equilibrium phase difference is insensitive to the direction of the the interface exchange field.
This result indicates that pairing states carrying the Josephson current are changed from opposite-spin components [ESE, OSO, ETO ($S=0$), OTE ($S=0$)] to equal-spin ones [ETO ($S=1$) and OTE ($S=1$)] between $L=12$ and $L=30$.
From this result, a short junction (in our setting, $L<20$) is considered to behave as an ordinary S/insulator/S junction where the chiral edge mode doesn't work as a conducting channel.

\begin{figure}
 \centering
 \includegraphics[width=88mm]{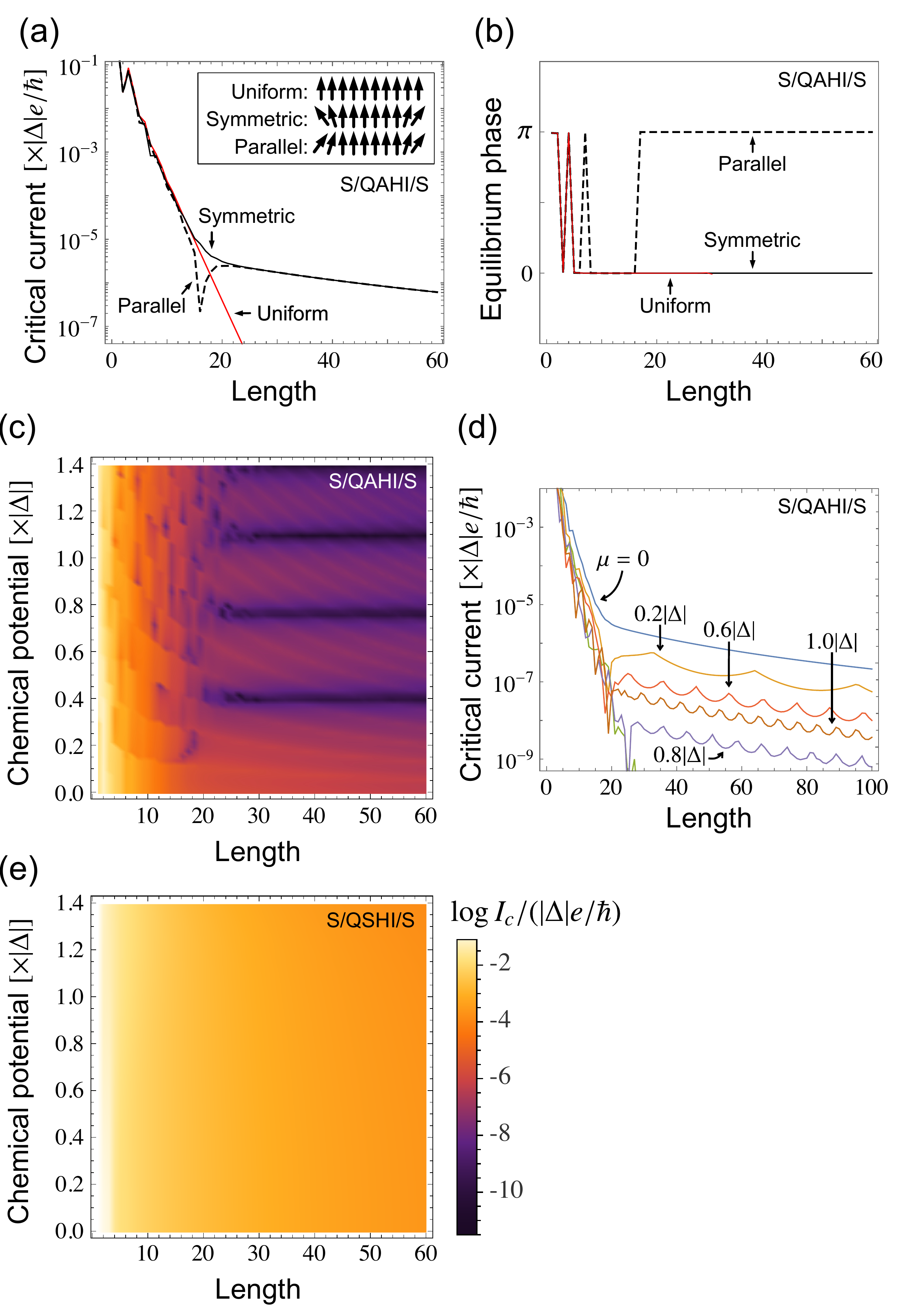}
 \caption{Length dependence of (a) the critical current and (b) the equilibrium phase difference is shown for the chemical potential $\mu=0$. The solid and dashed lines are for the exchange field of the left interface being parallel (the azimuth angle $\phi_\text{L}=0$) and symmetric ($\phi_\text{L}=\pi$) to that of the right interface, respectively, and the red line is for the case where the exchange field is uniform. The inset in (a) shows the uniform, symmetric and parallel configurations of the exchange field. The length and chemical-potential dependence of the critical current in (c) a S/QAHI/S junction with the symmetric configuration and in (e) a S/QSHI/S junction are shown in log scale. For a specific value of the chemical potential $\mu/|\Delta|=0,0.2,0.6,0.8,$ and $1$, the length dependence of a S/QAHI/S junction is shown in (d). The critical current $I_c$ in (c) and (e) is measured in unit of $|\Delta|e/\hbar$. 
 \label{fig:currentlength}}
\end{figure}

The QAHI length dependence of the critical current and the equilibrium phase difference is shown in Fig.~\ref{fig:currentlength} (a) and (b), respectively.
Around a length $L\sim 20$, an exponential decay implying that the QAHI behaves as an insulator is changed to an algebraic decay implying that the chiral edge mode carries the Josephson current.
This crossover can also be seen from the equilibrium phase in Fig.~\ref{fig:currentlength} (b), where a short junction is insensitive to the relative direction of the exchange field between two interfaces while the equilibrium phase in a long junction is changed from $\pi$ to 0 by rotating the azimuth angle of the left interface by $\pi$ (from parallel to symmetric configuration).
The critical current for the parallel configuration shows a dip around $L=20$ where the transition from 0 to $\pi$ junction occurs.
When the exchange field is uniform throughout the QAHI, triplet pairing states are not induced and thus the critical current decays exponentially even in a long junction ($L>20$) [the red curve in Fig.~\ref{fig:currentlength} (a)].

The length and chemical-potential dependence of the critical current is shown in Fig.~\ref{fig:currentlength} (c) and (d) for the S/QAHI/S junction and (e) for the S/QSHI/S junction, respectively.
In contrast to the S/QSHI/S junction, a periodic wavy pattern of a S/QAHI/S junction at $\mu>0$ and $L>20$ is attributed to the finite-size effect, where the critical current shows a cusp when the chemical potential $\mu$ agrees with a discrete spectrum of the chiral edge mode [Fig.~\ref{fig:spectralflow} (a)].
Notice that, the periodicity of the cusp pattern in Fig.~\ref{fig:currentlength} (d) is considered to represent the oscillation corresponding to the momentum $k_F$ (not $2k_F$) as a function of the perimeter of the QAHI.
The current phase relation in the S/QSHI/S junction is not sinusoidal due to the discrete ingap spectrum sensitive to the phase difference [Fig.~\ref{fig:spectralflow} (b)].

\subsection{Disorder effect}
\label{sec:disordereffect}

\begin{figure}
 \centering
 \includegraphics[width=84mm]{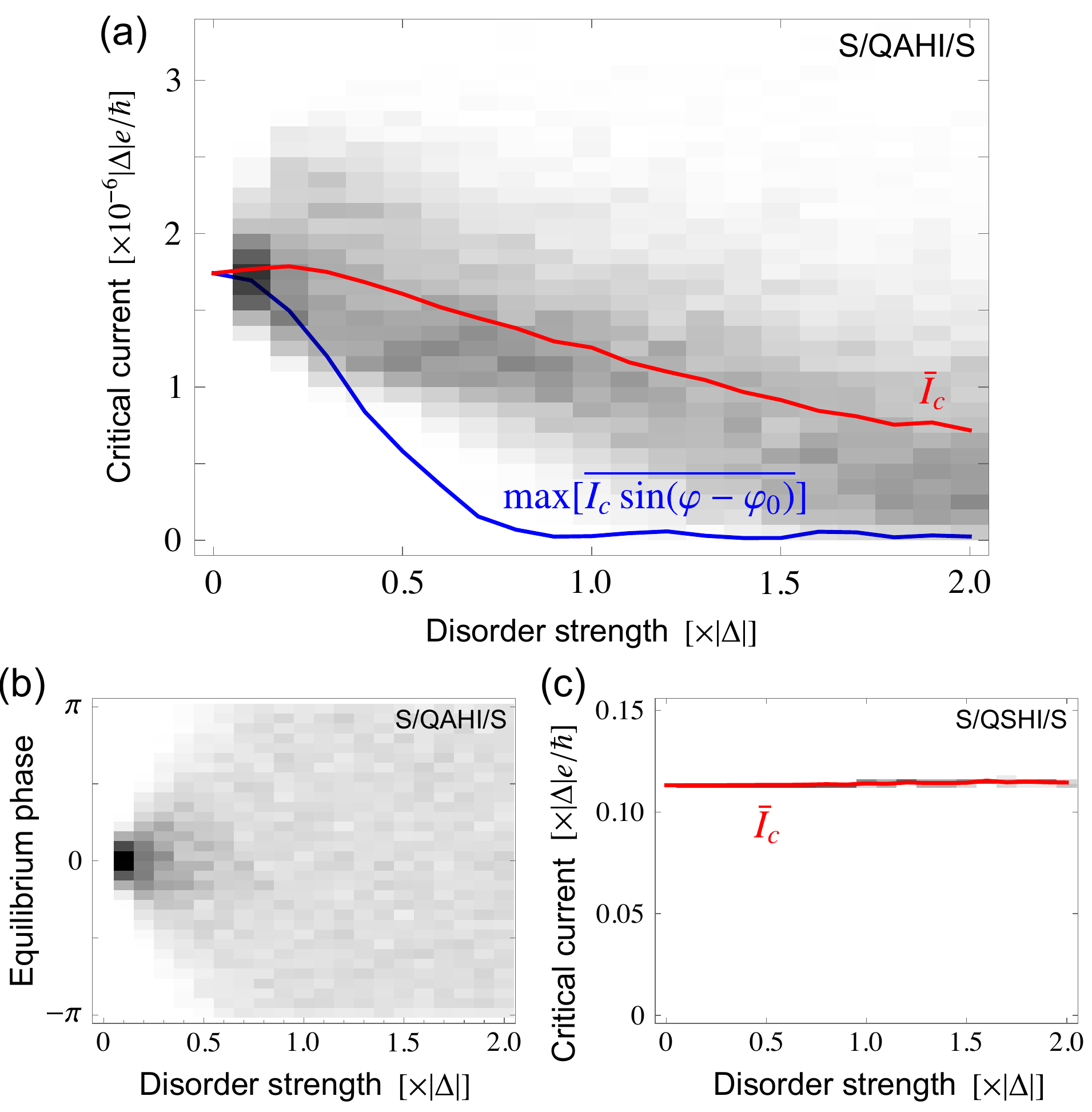}
 \caption{Disorder effect of (a) the critical current and (b) the equilibrium phase difference in a S/QAHI/S junction and that of (c) the critical current of a S/QSHI/S junction are shown.
 The gray scale of each rectangular region indicates the number of disorder configurations.
 The red curves in (a) and (c) represent the average of the critical current over 1000 and 10 disorder configurations, respectively, and the blue curve in (a) represents the critical current of the configuration-averaged current-phase relation.
 \label{fig:currendisorder}}
\end{figure}

Finally, we consider the disorder effect on the DC Josephson effect.
Here, we consider on-site potential disorder given in the Nambu space by
\begin{align}
 [\mathcal{H}_\text{disorder}]_{\bm{r}\bm{r}}
 =
 \begin{pmatrix}
  \delta\epsilon_{\bm{r}} & 0 \\
  0 & -\delta\epsilon_{\bm{r}}
 \end{pmatrix},
\end{align}
where $\bm{r}$ is the position in the QAHI$, \delta\epsilon_{\bm{r}}$ is uniformly distributed within $[-W,W]$, and each matrix element implicitly accompanies the identity matrices of the spin and orbital degrees of freedom. 
The critical current and the equilibrium phase difference in the presence of disorder of the strength $W$ are shown in Fig.~\ref{fig:currendisorder} (a) and (b), respectively.
It is known that within a disorder strength $W\in[0,2|\Delta|]$ studied here, the bulk is still in a QAHI phase and thus the chiral edge mode persists.
The results indicates that the both critical current $I_c$ and the equilibrium phase difference $\varphi_0$ are sensitive to disorder configurations, which can also be seen in rapidly decaying critical current of the configuration-averaged current-phase relation $\text{max}[\overline{I_c\sin(\varphi-\varphi_0)}]$ [the blue curve in Fig.~\ref{fig:currendisorder} (a)], while the average of the critical current $\bar{I}_c$ is less sensitive to disorder [the red curve in Fig.~\ref{fig:currendisorder} (a)].

Although disorder is an irrelevant perturbation to the chiral edge mode, it can change local parameters such as the Fermi velocity and the chemical potential.
When the chemical potential has asymmetry between the top and bottom boundaries, the equilibrium phase difference can be shifted \cite{sakurai17}.
This property will be studied analytically in Sec.~\ref{sec:analytical_asymmetry}.
It will be shown that the configuration-dependent equilibrium phase difference is attributed to asymmetry of the potential on top and bottom boundaries.
Notice that in order for the Josephson current through robust edge modes to be sensitive to disorder configurations, it is essential that the chiral edge mode extends across the boundaries [Fig.~\ref{fig:spectralflow}(c)], since the shift of the equilibrium phase difference is resulting from the interference of two conduction modes of the Josephson current: one is by an electron on the bottom boundary and a hole on the top boundary and the other is by a similar mode with the electron and hole inverted.
On the other hand, less sensitive averaged critical current [the red curve in Fig.~\ref{fig:currendisorder} (a)] is considered to be resulting from the robustness of the chiral edge mode (the critical current through a diffusive normal metal typically decays as $I_c\propto W^{-4}$, since when the Thouless energy $E_\text{Th}$ is smaller than the pair potential $|\Delta|$, the critical current obeys $I_cR_\text{N}\sim E_\text{Th}$ \cite{belzig99}, where the normal-metal resistance and the Thouless energy are estimated as $R_\text{N}\propto\tau^{-1}$ and $E_\text{Th}\propto\tau$, and the relaxation time as $\tau\propto W^{-2}$ by the Born approximation \cite{lee85}).
While the averaged critical current decreases as the disorder becomes stronger within $W\in[0,2|\Delta|(=t_s)]$, it increases as the disorder strength approaches the critical value $W\sim 4t_s$, over which the bulk becomes a (non-topological) Anderson insulator.

Fig.~\ref{fig:currendisorder} (c) shows the disorder effect of the critical current of the S/QSHI/S junction. 
Since the Andreev reflection in the S/QSHI/S junction occurs locally and thus there is no interference, the phase difference at an equilibrium is always 0. Also the critical current is insensitive to both disorder strength and disorder configurations.

\section{A chiral edge model for analytical calculations}
\label{sec:chiraledgemodel}

\begin{figure}
 \centering
 \includegraphics[width=84mm]{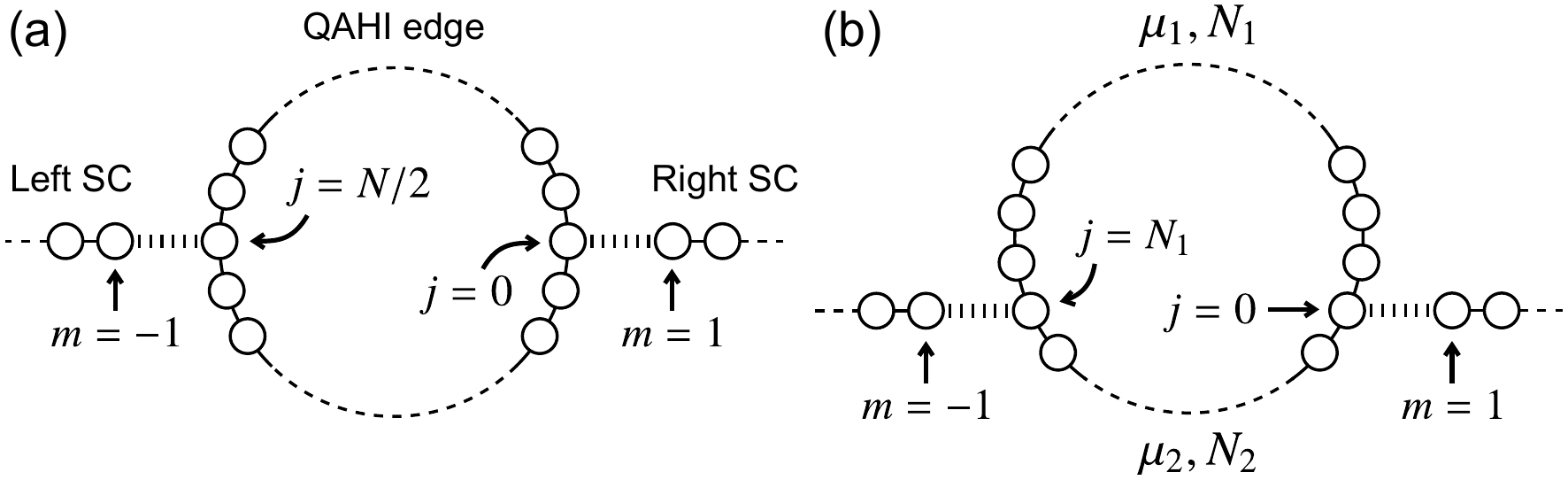}
 \caption{(a) Symmetric and (b) asymmetric lattice geometry of a analytical model of a QAHI edge connected to two superconductors.
 \label{fig:analytical}}
\end{figure}

In the following, the numerical results in the previous sections, that is, the edge-induced pairing states and the DC Josephson effect through a spin-polarized QAHI are studied analytically based on the equilibrium Green's function.
In this section, the analytical model of a Josephson junction is defined on a one-dimensional lattice [Fig.~\ref{fig:analytical} (a)].

The low-energy properties of the QAHI is modeled by the gapless edge mode: a spin-polarized chiral linear-dispersion model on a closed one-dimensional lattice defined by 
\begin{align}
 H_\text{edge}
 =
 \sum_k
 c_{k\uparrow}^{\dagger}(vk-\mu)
 c_{k\uparrow}. 
 \label{eq:hamiltonian_chiraledge}
\end{align}
Here, although the momentum $k=2\pi s/N$ for the lattice of $N$ sites should be defined in the Brillouin zone $[-\pi,\pi]$, this condition is relaxed to $k=2\pi s/N\,(s\in[-\infty,\infty])$ to remove the ambiguity \cite{ma93}.

The Josephson junction is modeled by the chiral edge model (\ref{eq:hamiltonian_chiraledge}) coupled with two $s$-wave superconductors on a one-dimensional semi-infinite lattice [Fig.~\ref{fig:analytical} (a)].
The Hamiltonian of the two superconductors are given by
\begin{align}
 &H_\text{SC,R}
 =
 -t_\text{SC}\sum_{m\ge 1,\sigma}
 a_{m+1\sigma}^{\dagger}a_{m\sigma}
 +
 \Delta_\text{R}\sum_{m\ge 1} a_{m\uparrow}^{\dagger}a_{m\downarrow}^{\dagger}
 +
 \text{h.c.}, 
 \label{eq:hamiltonian_rightsuperconductor}\\
 &H_\text{SC,L}
 =
 -t_\text{SC}\sum_{m\le -1,\sigma}
 a_{m-1\sigma}^{\dagger}a_{m\sigma}
 +
 \Delta_\text{L}\sum_{m\le -1} a_{m\uparrow}^{\dagger}a_{m\downarrow}^{\dagger}
 +
 \text{h.c.},
 \label{eq:hamiltonian_leftsuperconductor}
\end{align}
where the right (left) superconductor is defined on a lattice labeled by $m\ge 1 $ ($m\le -1$).
Notice that the annihilation and creation operators of the electrons in the chiral edge (the two superconductors) are denoted by $c$ and $c^{\dagger}$ ($a$ and $a^{\dagger}$).

Let the number of the lattice sites $N$ be an even integer and let the length of two paths (clockwise and anti-clockwise) connecting two contact points on the chiral edge be the same.
We assign $j=0$ ($j=N/2$) to the coordinate of the right (left) contact point.
The tunneling Hamiltonian between the chiral edge and the two superconductors is given by
\begin{align}
 H_\text{tunnel,R}
 &=
 \begin{pmatrix}
  a_{1\uparrow}^{\dagger} & a_{1\downarrow}^\dagger
 \end{pmatrix}
 T_\text{R}
 \begin{pmatrix}
  c_{0\uparrow} \\
  c_{0\downarrow}
 \end{pmatrix}
 +
 \text{h.c.}, \\
 H_\text{tunnel,L}
 &=
 \begin{pmatrix}
  a_{-1\uparrow}^{\dagger} & a_{-1\downarrow}^\dagger
 \end{pmatrix}
 T_\text{L}
 \begin{pmatrix}
  c_{N/2\uparrow} \\
  c_{N/2\downarrow}
 \end{pmatrix}
 +
 \text{h.c.},
\end{align}
where $c_{j\sigma}\,(\sigma=\uparrow,\downarrow)$ in the above expression is the Fourier transform of $c_{k\sigma}$ in (\ref{eq:hamiltonian_chiraledge}).
The tunneling matrix defined by
\begin{align}
 T_\text{R(L)}
 =
 -t_\text{tunnel}
 R_\text{R(L)}
 \left[
 \frac{1+\sigma^z}{2}
 +
 \tau
 \frac{1-\sigma^z}{2}
 \right] 
 R_\text{R(L)}^{-1},
 \label{eq:tunnelingmatrix}
\end{align}
 represents spin-filtered tunneling resulting from a tilted spin-polarization axis which works equivalently as the interface layers in the lattice model \cite{beri09}.
The polarization axis is measured by the polar angle $\theta_\text{R(L)}$ and the azimuth angle $\phi_\text{R(L)}$, and they are contained in the tunneling matrix through the rotation matrix $R_\text{R(L)}=R(\theta_\text{R(L)},\phi_\text{R(L)})=e^{-i\phi_\text{R(L)}\sigma^z/2}e^{-i\theta_\text{R(L)}\sigma^y/2}$ of spin 1/2.
Notice that the tunneling matrix contains tunneling of both majority and minority spins with respect to the polarization axis, the latter of which is resulting from the evanescent mode decaying exponentially by the thickness of the interface layer, and the ratio of the amplitude of two tunneling is denoted by $\tau (<1)$ \cite{beri09}.

\section{Analytical results}
\label{sec:analytical}

First of all, we review the Green's function of the chiral edge mode and the superconductors in the absence of the tunneling between them, the derivation of which is explained in Appendix \ref{sec:unperturbedgreensfunction}.

The Matsubara Green's function of the chiral edge electron and hole between two points $j$ and $j'$ is given, respectively, as \cite{ma93}
\begin{align}
 [g^\text{edge}_{jj'}(i\omega_n)]_{11}
 =&
 \frac{1}{2v}
 \frac{e^{-i(i\omega_n+\mu)(\text{sgn}[j-j']N/2-j+j')/v}}{\sin[N(i\omega_n+\mu)/2v]}, \\
 [g^\text{edge}_{jj'}(i\omega_n)]_{33}
 =&
 \frac{1}{2v}
 \frac{e^{-i(i\omega_n-\mu)(\text{sgn}[j-j']N/2-j+j')/v}}{\sin[N(i\omega_n-\mu)/2v]}.
\end{align}
On the other hand, the Green's function of the spin-down electron and hole that do not propagate along the chiral edge is also necessary in the following argument.
Here, the energy band of the spin-down electrons and holes is assumed to be dispersionless with an energy gap $V$, that is, 
$
H_{\downarrow}
 =
 \sum_k
 V
 c_{k\downarrow}^{\dagger}
 c_{k\downarrow}
$, 
and the resulting Green's function is, respectively,
\begin{align}
 [g^\text{edge}_{jj'}(i\omega_n)]_{22}
 &=
 \frac{1}{N}
 \sum_k
 e^{ik(j-j')}
 \frac{1}{i\omega_n-V}
 =
 \frac{\delta_{jj'}}{i\omega_n-V},\\
 [g^\text{edge}_{jj'}(i\omega_n)]_{44}
 &=
 \frac{\delta_{jj'}}{i\omega_n+V}.
\end{align}

The Green's function of the superconductors in a semi-infinite lattice is derived by using the boundary Green's function method \cite{arrachea09,zazunov16}.
In the wide-band limit ($\omega_n,|\Delta|\ll t_\text{SC}$), the boundary Green's function between the leftmost (rightmost) site of the right (left) superconductor is, respectively,
\begin{align}
 &g^\text{SC,R}_{11}(i\omega_n)
 \simeq
 -\frac{i\omega_n+\tilde{\Delta}_\text{R}}{t_\text{SC}(\omega_n^2+|\Delta_\text{R}|^2)^{1/2}},
 \label{eq:boundarygreensfunction_right}\\
 &g^\text{SC,L}_{-1-1}(i\omega_n)
 \simeq
 -\frac{i\omega_n+\tilde{\Delta}_\text{L}}{t_\text{SC}(\omega_n^2+|\Delta_\text{L}|^2)^{1/2}},
 \label{eq:boundarygreensfunction_left}
\end{align}
where $\tilde{\Delta}_\text{R(L)}=-\text{Re}[\Delta_\text{R(L)}]\sigma^y\tau^y-\text{Im}[\Delta_\text{R(L)}]\sigma^y\tau^x$.

\subsection{Anomalous Green's function}

First, consider the anomalous Green's function between two sites $j,j'$ on the chiral edge. 
When $j$ and $j'$ are away from the contact points, only spin-up components propagate. 
The leading order terms of the anomalous Green's function representing the equal-spin Andreev reflection in the perturbation series with respect to the tunneling amplitude $t_\text{tunnel}$ are given by  
\begin{align}
 &[G_{jj'}^{(2)}(i\omega_n)]_{13} 
 =
 -\frac{i\omega_nC_\text{R} e^{i(\text{arg}[\Delta_\text{R}]-\phi_\text{R})}}
 {(\omega_n^2+|\Delta_\text{R}|^2)(\omega_n^2+V^2)} \notag\\
 &\times 
 [g^\text{edge}_{j0}(i\omega_n)]_{11}
 [g^\text{edge}_{0j'}(i\omega_n)]_{33}
 +
 (\text{R}\to\text{L},0\to N/2),
 \label{eq:nonlocalanomalouscorrelation_2}
\end{align} 
where
$
 C_\text{R/L}
 =
 t_\text{tunnel}^4|\Delta_\text{R/L}|V \tau(1-\tau^2)
 \sin\theta_\text{R/L}/
 t_\text{SC}^2
$ (see derivation in Appendix \ref{sec:appendix_anomalousgreensfunction}).
Some notices on (\ref{eq:nonlocalanomalouscorrelation_2}) are in order:
the anomalous Green's function depends on
(i) the tunneling of the evanescent minority spin ($\propto\tau$), 
(ii) the reflection of the spin-down electrons and holes from a superconductor back to the same superconductor ($\propto [g^\text{edge}_{00}(i\omega_n)]_{22}-[g^\text{edge}_{00}(i\omega_n)]_{44}=-2V/(\omega_n^2+V^2)$), and also 
(iii) the polar angle of the tunneling matrix $T_\text{R/L}$ through $\sin\theta_\text{R/L}$. 
These facts indicate that the equal-spin Andreev reflection is resulting from the combination of spin-flipping normal reflection and the opposite-spin (ordinary) Andreev reflection.
In addition, 
(iv) the anomalous Green's function vanishes in the limit $\omega_n\to 0$, which is consistent with vanishing Andreev reflection in the same limit of the Josephson junction through half metals \cite{beri09}.
(v) The azimuth angle $\phi_\text{R/L}$ works equivalently as the phase of the pair potential $\Delta_\text{R/L}$ \cite{braude07,eschrig07,asano07-2,eschrig08,beri09}.
(vi) Since 
\begin{align}
 &[g^\text{edge}_{j0}(i\omega_n)]_{11}
 [g^\text{edge}_{0j'}(i\omega_n)]_{33} \notag\\
 &=
 \frac{e^{-\omega_n(j-j')/v}e^{-i\mu(N-j-j')/v}}
 {4v^2\sin[N(i\omega_n+\mu)/2v]\sin[N(i\omega_n-\mu)/2v]},
\end{align}
the anomalous Green's function with a fixed interval $j-j'$ behaves as $e^{2i\mu j/v}$, which agrees with a numerically found $2k_F(=2\mu/v)$ oscillation in Sec.~\ref{sec:pairing}.

The ETO and OTE components can be directly derived by substituting $i\omega_n\to\pm(\omega+ i\delta)$ and using the definition (\ref{eq:etootedefinition}), part of which resulting from the right superconductor is given by
\begin{align}
 &F^\text{ETO/OTE}_{jj+a}(\omega)
 =
 -\frac{\omega C_\text{R} e^{i(\text{arg}[\Delta_\text{R}]-\phi_\text{R})}}
 {v^2(|\Delta_\text{R}|^2-\omega^2)(V^2-\omega^2)} \notag\\
 &\times
 \frac{\chi^\text{ETO/OTE}(a)e^{-i\mu(N-2j)/v}}
 {\sin[N(\omega+i\delta+\mu)/2v]\sin[N(\omega+i\delta-\mu)/2v]}.
 \label{eq:nonlocalanomalouscorrelation_etoote}
\end{align}
Here, the interval $a$ appears in the expression as $\chi^\text{ETO}(a)=ie^{i\mu a/v}\sin[\omega a/v]$ and $\chi^\text{OTE}(a)=e^{i\mu a/v}\cos[\omega a/v]$, which agree with the periodicity of the nonlocal anomalous correlation in Sec.~\ref{sec:nonlocalcorrelation}.
From the expression of $\chi^\text{ETO/OTE}(a)$, large OTE indicates small ETO and vice versa \cite{tanaka07}.
Both the ETO and OTE components have singularities at $\omega=2\pi v s/N\mp\mu (s\in\mathbb{Z})$, which corresponds to the spectra of the chiral edge electron (hole).

\subsection{DC Josephson effect}

Next, consider the Josephson current through the chiral edge mode (see derivation in Appendix \ref{sec:appendix_josephsoneffect}).
The Josephson current is an equilibrium current, and thus is derived from the equilibrium Green's function as
\begin{align}
 I
 &=
 \frac{ie}{2\beta\hbar}
 \sum_{\omega_n}
 \text{tr}\,
 \tau^z
 \left[
 \mathcal{T}_\text{R}
 G_{10}(i\omega_n)
 -
 \mathcal{T}_\text{L}
 G_{-1N/2}(i\omega_n)
 \right],
 \label{eq:josephsoncurrent_greensfunction}
\end{align}
where $\mathcal{T}_\text{R(L)}$ is defined in (\ref{eq:tunnelingmatrix44}).
Notice that $0$ and $N/2$ in the subscript of the Green's functions in (\ref{eq:josephsoncurrent_greensfunction}) are the coordinate of the right and left point contacts on the chiral edge, respectively, and $1$ ($-1$) is the leftmost (rightmost) site of the right (left) superconductor.

Intuitively, the Josephson current is carried by electrons moving in one direction and holes moving in the opposite direction, the number of which is balanced to be in an equilibrium by the Andreev reflection at two superconductors \cite{furusaki91}.
In our model, electrons on the bottom (top) boundary and holes on the top (bottom) boundaries are related by the equal-spin Andreev reflection by the right and left superconductors, which is described by the two terms in (\ref{eq:nonlocalanomalouscorrelation_2}) by assigning $j$ and $j'$ on the bottom and top (bottom and top) boundaries, respectively.
The leading-order contribution to the Josephson current is thus the combination of the leading contribution of the anomalous Green's function $[G_{N/2,0}^{(2)}(i\omega_n)]_{13(31)}$ given in (\ref{eq:nonlocalanomalouscorrelation_2}) and and its counterpart $[G_{0,N/2}^{(2)}(i\omega_n)]_{31(13)}$
The expression of the Josephson current is shown in (\ref{eq:josephsoncurrent_thirdorder}).

In the high-temperature limit ($2\pi v/N\ll \beta^{-1}$), the summation over the Matsubara frequency $\omega_n$ is approximated by the summation over $\omega_{0(1)}=\pm\pi\beta^{-1}$, which gives
\begin{align}
 I^{(3)}
 &\simeq
 \frac{e}{\hbar}
 \frac{4\beta^5C_\text{R}C_\text{L}}
 {v^2}e^{-N\pi/\beta v}
 \sin
 \delta\varphi\notag\\
 &\times
 \frac{\pi^2}{(\pi^2+\beta^2|\Delta_\text{R}|^2)(\pi^2+\beta^2|\Delta_\text{L}|^2)(\pi^2+\beta^2V^2)^2},
 \label{eq:josephsoncurrent_hightemperature}
\end{align}
where
\begin{align}
 \delta\varphi
 =
 \text{arg}\frac{\Delta_\text{L}}{\Delta_\text{R}}
 +\phi_\text{R}-\phi_\text{L}.
 \label{eq:phasedifference}
\end{align}
On the other hand, in the low-temperature limit ($2\pi v/N\gg \beta^{-1}$) and when the inter-level spacing of the chiral edge mode is much smaller than the pair potential and the gap of the minority spin ($2\pi v/N\ll|\Delta_{R/L}|, V$) the Josephson current is given by
\begin{align}
 I^{(3)}
 \simeq
 \frac{e}{\hbar}
 \frac{2\pi vC_\text{R}C_\text{L}}
 {3N^3V^4|\Delta_\text{R}\Delta_\text{L}|^2}
 F\left(\frac{N\mu}{2\pi v}\right)
 \sin
 \delta\varphi,
 \label{eq:josephsoncurrent_lowtemperature}
\end{align}
where $F(a)$ is a dimensionless periodic function [$F(a+1)=F(a)$] defined in (\ref{eq:periodicfunction}), and it gives the maximum value of 1 when the chemical potential lies exactly at the discrete spectrum of the chiral edge mode $\mu=2\pi v s/N (s\in\mathbb{Z})$ (see Fig.~\ref{fig:periodicfunction}).
The Josephson current is dependent on the relative azimuth angle $\phi_R-\phi_L$ of the tunneling matrix in addition to the phase difference of the pair potential of two superconductors, which shares the same property as the Josephson current through half metals \cite{braude07,eschrig07,asano07-2,eschrig08,beri09}.
The phase (\ref{eq:phasedifference}) is the consequence of the interference mentioned in Sec.~\ref{sec:disordereffect}.
The critical current decays as $1/N^3$, which agrees with \cite{vanostaay11}.

\subsection{Asymmetric geometry}
\label{sec:analytical_asymmetry}

Next, we consider a case where the geometry of the junction has asymmetry with respect to the permutation of the top and bottom boundaries [Fig.~\ref{fig:analytical} (b)] (see details in Appendix \ref{sec:appendix_aymmetric}).
A similar topic has been studied from the perspective of symmetry of a spin-unpolarized QAHI in \cite{sakurai17}.

Let the length and the chemical potential of the top (bottom) boundary be denoted by $N_1$ and $\mu_1$ ($N_2$ and $\mu_2$).
The plane wave $e^{ikx}$ is an eigenstate with the eigenenergy $vk-\mu_1$ ($vk-\mu_2$) on the top (bottom) boundary.
Eigenstates in the whole chiral edge are thus determined by the scattering problem, in which plane waves with the same eigenenergy are connected by boundary conditions.
The boundaries of the two regions are the two contact points with superconductors.
Let the perimeter of the chiral edge be denoted by $N (=N_1+N_2)$.
The eigenenergies are $\epsilon=2\pi vs/N-\tilde{\mu} (s\in\mathbb{Z})$, where $\tilde{\mu}=(\mu_1N_1+\mu_2N_2)/N$, and the corresponding momentum is  $k_{1s}=2\pi s/N+(\mu_1-\mu_2)N_2/Nv$ and $k_{2s}=2\pi s/N+(\mu_2-\mu_1)N_1/Nv$ on the top and bottom boundaries, respectively.
The unperturbed Green's function of the chiral edge without tunneling can be derived in the same way as before.
The resulting expression of the Josephson current is obtained from that for the symmetric case (\ref{eq:josephsoncurrent_thirdorder}), (\ref{eq:josephsoncurrent_hightemperature}), and (\ref{eq:josephsoncurrent_lowtemperature}) by changing the chemical potential as $\mu\to\tilde{\mu}$ and $\delta\varphi$ defined in (\ref{eq:phasedifference}) by
\begin{align}
 \delta\tilde{\varphi}
 =
 \delta\varphi
 -
 \frac{(N_1-N_2)\tilde{\mu}}{v}
 -
 \frac{2N_1N_2(\mu_1-\mu_2)}{Nv}.
 \label{eq:phasedifference2}
\end{align}
The dependence of the length difference appears in the second term of the right-hand side of (\ref{eq:phasedifference2}) and that of the chemical-potential difference in the third term.

According to (\ref{eq:phasedifference2}), even when the length of the two boundaries is the same ($N_1=N_2$), a chemical-potential difference of the interlevel spacing of the chiral edge mode ($\mu_1-\mu_2=2\pi v/N$) can change a 0 junction into a $\pi$ junction or vice versa. 
Therefore, although the expression in this section is applicable only to systems without disorder, it can be deduced that strong disorder-configuration dependence of the equilibrium phase in Sec.~\ref{sec:disordereffect} is the consequence of asymmetric profile of the disorder potential.

\section{Conclusion}
\label{sec:conclusion}

We have studied the Josephson effect through a spin-polarized QAHI between conventional $s$-wave superconductors focusing on pairing symmetry induced by the edge state and on the comparison with that through a QSHI.
The pairing states along the QAHI edge are equal-spin triplet, the combination of even and odd frequencies, nonlocally extended, and have a finite momentum (the Fulde-Ferrell state) due to the spin polarization, translation symmetry breaking, robustness, and the chirality.

Equal-spin triplet pairs carry the Josephson current when the length of the QAHI exceeds a crossover length, below which the Josephson current is carried by singlet pairs via the bulk.
The appearance of the equal-spin triplet pairs was examined by the dependence of the interface-exchange-field direction between the QAHI and two superconductors.
When the three directions of the bulk, left- and right-interface exchange field have a finite solid angle, the phase difference that minimizes the junction energy is neither 0 nor $\pi$, and changes continuously by $2\pi$ during the rotation of the left-interface exchange field by $2\pi$ around the bulk one.
The finite momentum of the ETO and OTE pairs was confirmed by a spatial oscillation of the corresponding anomalous Green's function, in which the periodicity coincides with twice the Fermi momentum $2k_F$.

Numerical calculations are qualitatively examined by a simple analytical model of a chiral fermion on a closed one-dimensional circle coupled to two superconductors on a semi-infinite lattice.
The anomalous Green's function and the Josephson current are perturbatively derived via the Matsubara Green's function up to the leading order in the tunneling amplitude between the QAHI and superconductors.
The analytical model explains a numerically discovered large fluctuation of the equilibrium phase difference in the presence of potential disorder from an asymmetric profile of the chemical potential.

\begin{acknowledgments}
This work was supported by the Japan Society for the Promotion of Science KAKENHI (Grant Nos. JP17K17604, JP20H01830, JP18H01176, and JP20H01857) and by CREST, Japan Science and Technology Agency (Grant No. JPMJCR18T2). Y.T. was also supported by Grant-in-Aid for Scientific Research A (KAKENHI Grant No. JP20H00131) and the JSPS Core-to-Core program Oxide Superspin International Network.
\end{acknowledgments}

\appendix

\section{Unperturbed Green's function}
\label{sec:unperturbedgreensfunction}

In this section, we review the derivation of the unperturbed Green's function of the chiral edge fermion \cite{ma93} and the $s$-wave superconductor on a semi-infinite space \cite{arrachea09,zazunov16}.

First, the Matsubara Green's function of the spin-up electron on the chiral edge mode (\ref{eq:hamiltonian_chiraledge}) with the Matsubara frequency $\omega_n=(2n+1)\pi T$ between two points $j$ and $j'$ is given by
\begin{align}
 [g^\text{edge}_{jj'}(i\omega_n)]_{11}
 =
 \frac{1}{N}
 \sum_k
 e^{ik(j-j')}
 \frac{1}{i\omega_n-vk+\mu}.
\end{align}
Here, the momentum $k$ is simplified to be unbounded since the energy levels away from the Fermi level would be less relevant \cite{ma93}.
The summation over $k=2\pi s/N (s\in\mathbb{Z})$ is rewritten by the complex integral as
\begin{align}
 [g^\text{edge}_{jj'}(i\omega_n)]_{11}
 =&
 \frac{1}{2\pi iN}
 \oint_C 
 dz
 \frac{\text{sgn}[j-j']}{e^{\text{sgn}[j-j']z}-1}
 \frac{e^{z(j-j')/N}}{i\omega_n+ivz/N+\mu} \notag\\
 =&
 \frac{1}{2v}
 \frac{e^{-i(i\omega_n+\mu)(\text{sgn}[j-j']N/2-j+j')/v}}{\sin[N(i\omega_n+\mu)/2v]},
 \label{eq:greensfunction_barechiraledge}
\end{align}
where $C$ is an infinitely large circle on the complex plane that avoids a pole at $z=iN(i\omega_n+\mu)/v$, and $\text{sgn}[j-j']$ is included to ensure the convergence.
The expression (\ref{eq:greensfunction_barechiraledge}) is consistent with the periodicity of the number of the lattice sites $N$, that is, for two points $j$ and $j'$ satisfying $j>j'$, the distance measured in the clockwise direction ($j-j'>0$) and that in the counter-clockwise directions ($j-j'-N<0$) give the same result as long as they are within $[-N,N]$.
The unperturbed Green's function of the spin-up hole is given by changing $\mu\to-\mu$ in (\ref{eq:greensfunction_barechiraledge}).

The unperturbed Green's function of a superconductor on a semi-infinite lattice (the boundary Green's function) is derived by solving the Dyson equation for the superconductor on the infinite lattice in the presence of an infinite potential at $m=0$ that excludes particles at this site \cite{arrachea09,zazunov16}.
The Hamiltonian of the superconductor in the infinite lattice is $H_\text{SC}=H_\text{SC,R}+H_\text{SC,L}+H_\text{SC,0}$, where $\Delta_\text{R}=\Delta_\text{L}=\Delta$ and $H_\text{SC,0}$ connects the two superconductors as
\begin{align}
 H_\text{SC,0}
 =
 -t_\text{SC}
 \sum_\sigma
 \sum_{m=-1,0}
 a_{m+1\sigma}^{\dagger}a_{m\sigma}
 +
 \Delta a_{0\uparrow}^{\dagger}a_{0\downarrow}^{\dagger}
 +
 \text{h.c.}.
\end{align}
The Green's function is derived in the presence of a potential, the Hamiltonian of which is given by
\begin{align}
 H_\text{potential}=U\sum_{\sigma}a_{0\sigma}^{\dagger}a_{0\sigma}.
 \label{eq:hamiltonian_hardwall}
\end{align}

By Fourier transform of $H_\text{SC}$,
the Matsubara Green's function of the superconductor in the infinite lattice is
\begin{align}
 g^\text{SC}_{mm'}(i\omega_n)
 =
 \frac{1}{M}
 \sum_k
 e^{ik(m-m')}
 \frac{1}{i\omega_n-\mathcal{H}_\text{SC}(k)},
\end{align}
where $M$ is the number of lattice sites. 
Here, $\mathcal{H}_\text{SC}(k)=-2t\cos k \tau^z-\text{Re}[\Delta_\text{R}]\sigma^y\tau^y-\text{Im}[\Delta_\text{R}]\sigma^y\tau^x$ is the $4\times 4$ Hamiltonian matrix of $H_\text{SC}$ on a basis $(a_{k\uparrow},a_{k\downarrow},a^{\dagger}_{-k\uparrow},a^{\dagger}_{-k\downarrow})$, and $\sigma(\tau)$ is the Pauli matrix for the spin (Nambu) space.
In the limit of $M\to\infty$,
\begin{align}
 &g^\text{SC}_{mm'}(i\omega_n)
 \to
 \frac{1}{2\pi}
 \int dk
 e^{ik(m-m')}
 \frac{1}{i\omega_n-\mathcal{H}_\text{SC}(k)} \notag\\
 &=
 \frac{1}{2\pi i}
 \oint_{C_0} dz\,
 z^{m-m'-1}
 \frac{i\omega_n-t(z+z^{-1})\tau^z+\tilde{\Delta}}
 {\omega_n^2+t^2(z+z^{-1})^2+|\Delta|^2},
\end{align}
where $C_0$ is the unit circle in the complex plane, and $\tilde{\Delta}=-\text{Re}[\Delta]\sigma^y\tau^y-\text{Im}[\Delta]\sigma^y\tau^x$.
Performing the integral, we obtain
\begin{align}
 &g^\text{SC}_{mm}
 =
 -
 \frac{1}{t_\text{SC}^2}
 \frac{i\omega_n+\tilde{\Delta}}{\lambda^{-2}-\lambda^2}, 
 \label{eq:greenfunction_superconductorinfinite1}\\
 &g^\text{SC}_{m+1m}
 =
 g^\text{SC}_{mm+1}
 =
 \frac{1}{t_\text{SC}}
 \frac{\tau^z}{\lambda^{-2}+1},
 \label{eq:greenfunction_superconductorinfinite2}
\end{align}
where $\lambda^2=[\omega_n^2+|\Delta|^2+2t^2-(\omega_n^2+|\Delta|^2)^{1/2}(\omega_n^2+|\Delta|^2+4t^2)^{1/2}]/2t^2$.

In the presence of the potential (\ref{eq:hamiltonian_hardwall}), the Dyson equation for the boundary Green's function is given by
\begin{align}
 g^\text{SC,b}_{mm'}(i\omega_n)
 =
 g^\text{SC}_{mm'}(i\omega_n)
 +
 g^\text{SC}_{m0}(i\omega_n)
 \mathcal{U}
 g^\text{SC,b}_{0m'}(i\omega_n),
 \label{eq:dysonequation_superconductorhardwall}
\end{align}
where $\mathcal{U}=U\tau^z$.
In the limit of an infinite potential height $U\to \infty$, the Green's function $g^\text{SC,b}_{0m'}(i\omega_n)$ vanishes while the product with $\mathcal{U}$ remains finite.
The Dyson equation for $\mathcal{U}g^\text{SC,b}_{0m}(i\omega_n)$ is simplified by taking the limit of $U\to\infty$ as
\begin{align}
 \mathcal{U}
 g^\text{SC,b}_{0m}(i\omega_n)
 &=
 [\mathcal{U}^{-1}-
 g^\text{SC}_{00}(i\omega_n)]^{-1}
 g^\text{SC}_{0m}(i\omega_n) \notag\\
 &\to
 -[g^\text{SC}_{00}(i\omega_n)]^{-1}
 g^\text{SC}_{0m}(i\omega_n).
\end{align}
Therefore, the boundary Green's function of the superconductor is given in terms of that in the infinite lattice by
\begin{align}
 g^\text{SC,b}_{mm'}(i\omega_n)
 =
 g^\text{SC}_{mm'}(i\omega_n)
 -
 g^\text{SC}_{m0}(i\omega_n)
 [g^\text{SC}_{00}(i\omega_n)]^{-1}
 g^\text{SC}_{0m'}(i\omega_n).
\end{align}
Specifically, by substituting (\ref{eq:greenfunction_superconductorinfinite1}) and (\ref{eq:greenfunction_superconductorinfinite2}), the boundary Green's function between the leftmost (rightmost) site of the right (left) superconductor is given in the wide-band limit ($t\gg \omega_n,|\Delta_\text{R}|$) by (\ref{eq:boundarygreensfunction_right}) [(\ref{eq:boundarygreensfunction_left})].
Notice that since conventional superconductors are not topological, the boundary Green's function (\ref{eq:boundarygreensfunction_right}) and (\ref{eq:boundarygreensfunction_left}) are simply twice that in a uniform system (\ref{eq:greenfunction_superconductorinfinite1}) in this limit.

\section{Anomalous edge correlation}
\label{sec:appendix_anomalousgreensfunction}

In the presence of tunneling between the chiral edge mode and the two superconductors, the Dyson equation of the Green's function of the chiral edge mode is
\begin{align}
 G_{jj'}(i\omega_n)
 =& 
 g^\text{edge}_{jj'}(i\omega_n)
 +
 g^\text{edge}_{j0}(i\omega_n)
 \Sigma^\text{R}(i\omega_n)
 G_{0j'}(i\omega_n) \notag\\
 &+
 g^\text{edge}_{jN/2}(i\omega_n)
 \Sigma^\text{L}(i\omega_n)
 G_{N/2\,j'}(i\omega_n).
 \label{eq:dysonequation_chiraledge}
\end{align}
Here, by introducing $4\times 4$ tunneling matrix
\begin{align}
 \mathcal{T}_\text{R(L)}
 =
 \begin{pmatrix}
  T_\text{R(L)} & 0 \\
  0 & -T_\text{R(L)}^{\ast}
 \end{pmatrix},
 \label{eq:tunnelingmatrix44}
\end{align}
and using the boundary Green's function (\ref{eq:boundarygreensfunction_right}) and (\ref{eq:boundarygreensfunction_left}) of the two superconductors, the self energy is given by
\begin{align}
 &\Sigma^\text{R(L)}(i\omega_n)
 =
 \mathcal{T}_\text{R(L)}g^\text{SC,R(L)}_{11(-1-1)}(i\omega)\mathcal{T}_\text{R(L)} \notag\\
 &=
 -\frac{t_\text{tunnel}^2}{t_\text{SC}(\omega_n^2+|\Delta_\text{R(L)}|^2)^{1/2}} 
 \begin{pmatrix}
  i\omega_nP_\text{R(L)} & -i\tau\Delta_\text{R(L)}\sigma^y \\
  i\tau\Delta_\text{R(L)}^{\ast}\sigma^y & i\omega_nP_\text{R(L)}^{\ast}
 \end{pmatrix}.
 \label{eq:selfenergy}
\end{align}
By using projection operators $P_\pm(\theta_\text{R},\phi_\text{R})$ defined by
$
 T_\text{R}
 =
 -t_\text{tunnel}
 \left[
 P_+(\theta_\text{R},\phi_\text{R})
 +
 \tau
 P_-(\theta_\text{R},\phi_\text{R})
 \right]
$ [see also the definition of the tunneling matrix (\ref{eq:tunnelingmatrix})],
we obtain $P_\text{R(L)}=P_+(\theta_\text{R(L)},\phi_\text{R(L)})+\tau^2 P_-(\theta_\text{R(L)},\phi_\text{R(L)})$.

Since only the spin-up electrons and holes can propagate along the edge, the only finite components of the anomalous Green's function are the $(1,3)$ and $(3,1)$ components.
According to (\ref{eq:selfenergy}) the $(1,3)$-component of the self energy $\Sigma^\text{R(L)}(i\omega_n)$ is absent.
The leading order contribution in perturbation expansion appears in the second order in the self energy, and is given by
\begin{align}
 [G_{jj'}^{(2)}(i\omega_n)]_{13} 
 =&
 [g^\text{edge}_{j0}(i\omega_n)]_{11}
 [g^\text{edge}_{0j'}(i\omega_n)]_{33} \notag\\
 &\times
 [\Sigma^\text{R}(i\omega_n)
 g^\text{edge}_{00}(i\omega_n)
 \Sigma^\text{R}(i\omega_n)]_{13} \notag\\
 &+(\text{R}\to\text{L},0\to N/2).
 \label{eq:anomalouscorrelation_appendix}
\end{align}
Substituting (\ref{eq:selfenergy}), the anomalous Green's function is given by (\ref{eq:nonlocalanomalouscorrelation_2}).

\section{Josephson current}
\label{sec:appendix_josephsoneffect}

\subsection{Formula}

First, the formula of the Josephson current in terms of the Matsubara Green's function is reviewed.
Let the electric current flowing from the chiral edge to the right superconductor be denoted by $I_\text{R}$, and that from the left superconductor to the chiral edge by $I_\text{L}$.
We define the Josephson current flowing from the left to right superconductor by the average of the two current $I=(I_\text{R}+ I_\text{R})/2$.
The current operator through the right contact is determined by the change of the total electric charge $N_\text{R}$ of the right superconductor by the tunneling operator as
$
 \hat{I}_\text{R}
 =
 ie[N_\text{R},H_\text{tunnel,R}]/\hbar
$.
The Josephson current is given by 
\begin{align}
 I
 &=
 -\frac{e}{2\hbar}
 \text{tr}
 \left[
 \tau^z\mathcal{T}_\text{R}
 G^{-+}_{10}(t,t)
 -
 \tau^z\mathcal{T}_\text{L}
 G^{-+}_{-1\,N/2}(t,t)
 \right].
\end{align}
Here, 
$
 [G^{-+}_{mj}(t,t')]^{\alpha\beta}=-i\langle \Psi^{\alpha}_m(t)\Phi^{\dagger\beta}_j(t')\rangle
$
is the greater Green's function between a site $m$ on the superconductors and a site $j$ on the chiral edge, where $\Psi_m=(a_{m\uparrow}\,a_{m\downarrow}\,a_{m\uparrow}^{\dagger}\,a_{m\downarrow}^{\dagger})^T$ and $\Phi_j=(c_{j\uparrow}\,c_{j\downarrow}\,c_{j\uparrow}^{\dagger}\,c_{j\downarrow}^{\dagger})^T$.
In an equilibrium, the Fourier transform of the greater Green's function is related to the equilibrium Green's function through $G^{-+}(\omega)=(1-f(\omega))[G^R(\omega)-G^A(\omega)]$, where $f(\omega)$ is the Fermi distribution function.
Since the retarded (advanced) Green's function has poles in the lower-(upper-)half of the complex plane, the integral over the real frequency is closed by connecting the end points through a semicircle on the upper-(lower-)half plane, the integration path of which is indicated by $C_+$ ($C_-$).
Since poles inside the path are those of the Fermi distribution function at $i\omega_n=(2n+1)\pi i\beta^{-1}$, where $n\ge 0$ for $C_+$ and $n<0$ for $C_-$, we obtain
\begin{align}
 &G^{-+}(t,t)
 =
 \frac{1}{2\pi}
 \int d\omega\,
 (1-f(\omega))
 \left[
 G^{R}(\omega)
 -
 G^{A}(\omega)
 \right] \notag\\
 &=
 i
 \left[
 \frac{1}{2\pi i}
 \oint_{C_+} d\omega\,
 G^{R}(\omega)
 -
 \frac{1}{2\pi i}
 \oint_{C_-} d\omega\,
 G^{A}(\omega)
 \right]
 (1-f(\omega)) \notag\\
 &=
 -i\beta^{-1}
 \sum_{\omega_n}G(i\omega_n).
\end{align}
Then the Josephson current is given by
\begin{align}
 I
 &=
 \frac{ie}{2\beta\hbar}
 \sum_{\omega_n}
 \text{tr}
 \left[
 \tau^z\mathcal{T}_\text{R}
 G_{10}(i\omega_n)
 -
 \tau^z\mathcal{T}_\text{L}
 G_{-1\,N/2}(i\omega_n)
 \right].
\end{align}

\subsection{Perturbative expansion}

The Dyson equation for the Green's function along the chiral edge is given in (\ref{eq:dysonequation_chiraledge}), and that between the chiral edge and superconductors by
\begin{align}
 &\mathcal{T}^\text{R}G_{10}(i\omega_n)
 = 
 \Sigma^\text{R}(i\omega_n)
 G_{00}(i\omega_n), 
 \label{eq:dysonequation_chiraledgerightsuperconductor}\\
 &\mathcal{T}^\text{L}
 G_{-1N/2}(i\omega_n)
 = 
 \Sigma^\text{L}(i\omega_n)
 G_{N/2\,N/2}(i\omega_n).
 \label{eq:dysonequation_chiraledgeleftsuperconductor}
\end{align}
The leading non-vanishing contribution to the Josephson current is the combination of the leading contribution to the Andreev reflection described by (\ref{eq:anomalouscorrelation_appendix}) at both superconductors, which is given by the anomalous components of  
$
\Sigma^\text{R}(i\omega_n)
 g^\text{edge}_{00}(i\omega_n)
 \Sigma^\text{R}(i\omega_n)
$.
Therefore, the leading order of (\ref{eq:dysonequation_chiraledgerightsuperconductor}) and (\ref{eq:dysonequation_chiraledgeleftsuperconductor}) that has a finite contribution to the Josephson current is
\begin{align}
 &\mathcal{T}^\text{R}G_{10}
 \simeq
 [\Sigma^\text{R}
 g^\text{edge}_{00}
 \Sigma^\text{R}]
 g^\text{edge}_{0N/2}
 [\Sigma^\text{L}
 g^\text{edge}_{N/2N/2}
 \Sigma^\text{L}]
 g^\text{edge}_{N/2\,0} \notag\\
 &\qquad+
 \Sigma^\text{R}
 g^\text{edge}_{0\,N/2}
 [\Sigma^\text{L}
 g^\text{edge}_{N/2\,N/2}
 \Sigma^\text{L}]
 g^\text{edge}_{N/2\,0}
 \Sigma^\text{R}
 g^\text{edge}_{00},
 \label{eq:dysonequation_chiraledgerightsuperconductor2}\\
 &\mathcal{T}^\text{L}
 G_{-1N/2}
 \simeq
 \Sigma^\text{L}
 g^\text{edge}_{N/2\,0}
 [\Sigma^\text{R}
 g^\text{edge}_{00}
 \Sigma^\text{R}]
 g^\text{edge}_{0\,N/2}
 \Sigma^\text{L}
 g^\text{edge}_{N/2\,N/2} \notag\\
 &\qquad+
 [\Sigma^\text{L}
 g^\text{edge}_{N/2\,N/2}
 \Sigma^\text{L}]
 g^\text{edge}_{N/2\,0}
 [\Sigma^\text{R}
 g^\text{edge}_{00}
 \Sigma^\text{R}]
 g^\text{edge}_{0\,N/2},
 \label{eq:dysonequation_chiraledgeleftsuperconductor2}
\end{align}
where the Matsubara frequency in the argument is omitted.
The leading contribution to the Josephson current is given by
\begin{align}
 &I^{(3)}
 =
 -\frac{ie}{\hbar}
 \frac{t_\text{tunnel}^8V^2|\Delta_\text{R}\Delta_\text{L}|}
 {\beta t_\text{SC}^4}
 \tau^2(1-\tau^2)^2\sin\theta_\text{R}\sin\theta_\text{L}
 \sum_{\omega_n}\notag\\
 &\times
 \left(
 [g^\text{edge}_{0\,N/2}]_{33}
 [g^\text{edge}_{N/2\,0}]_{11}
 e^{-i\delta\varphi}
 -
 [g^\text{edge}_{0\,N/2}]_{11}
 [g^\text{edge}_{N/2\,0}]_{33}
 e^{i\delta\varphi}
 \right) \notag\\
 &\times
 \frac{\omega_n^2}{(\omega_n^2+|\Delta_\text{R}|^2)(\omega_n^2+|\Delta_\text{L}|^2)(\omega_n^2+V^2)^2},
 \label{eq:josephsoncurrent_thirdorder}
\end{align}
where $\delta\varphi$ is defined in (\ref{eq:phasedifference}).
The two terms in the parenthesis of the RHS of (\ref{eq:josephsoncurrent_thirdorder}) represent the interference between two current-carrying channels.

\begin{figure}
 \centering
 \includegraphics[width=60mm]{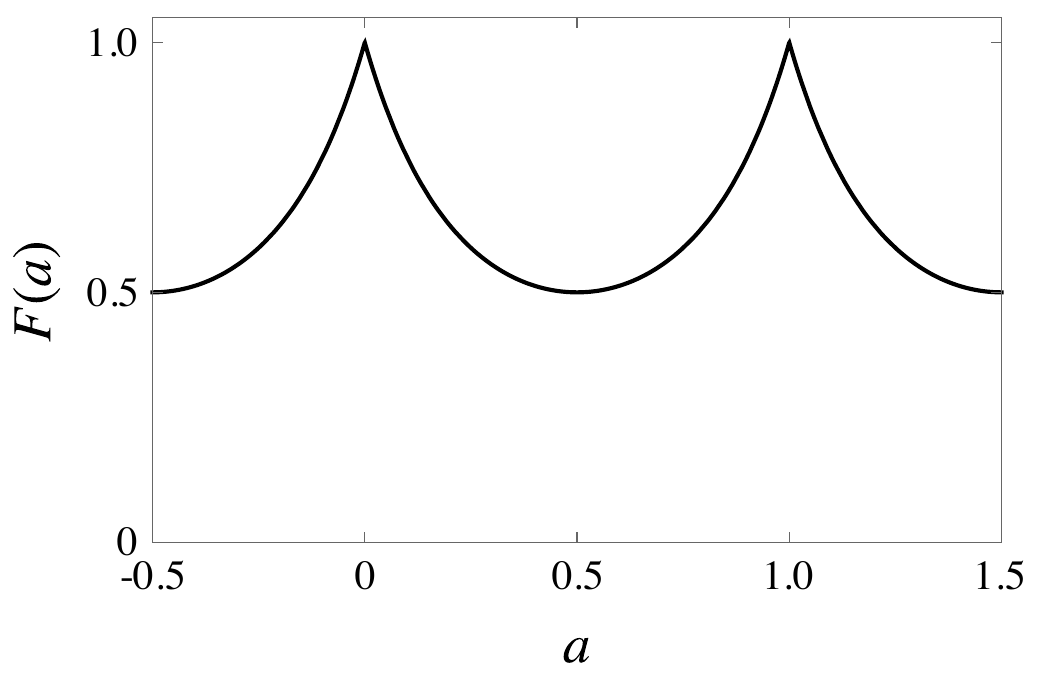}
 \caption{A periodic function $F(a)$ used in the low-temperature limit of the Josephson current.
 \label{fig:periodicfunction}}
\end{figure}

In the limit of low-temperature, the summation over the Matsubara frequency is replaced by the integral $\sum_{\omega_n}\to (\beta/2\pi)\int d\omega$.
Since 
\begin{align}
 &[g^\text{edge}_{0\,N/2}]_{33}
 [g^\text{edge}_{N/2\,0}]_{11}
 =
 [g^\text{edge}_{0\,N/2}]_{11}
 [g^\text{edge}_{N/2\,0}]_{33} \notag\\
 &=
 \frac{1}{2v^2}
 \frac{1}{
 \cos [N\mu/v]-\cosh[N\omega_n/v]}
\end{align}
decays exponentially by an energy of $2\pi v/N$ as a function of the Matsubara frequency $\omega_n$, the denominator of the third line of (\ref{eq:josephsoncurrent_thirdorder}) is approximated, by assuming that $2\pi v/N\ll|\Delta_\text{R(L)}|,V$, as $(\omega_n^2+|\Delta_\text{R}|^2)(\omega_n^2+|\Delta_\text{L}|^2)(\omega_n^2+V^2)^2\simeq|\Delta_\text{R}\Delta_\text{L}|^2V^4$. 
Then, the Josephson current is given by (\ref{eq:josephsoncurrent_lowtemperature}), where
\begin{align}
 F(a)
 =
 -\frac{3}{4\pi^2}\int dz
 \frac{z^2}{\cos 2\pi a-\cosh z}
 \label{eq:periodicfunction}
\end{align}
is a periodic function $F(a+1)=F(a)$ that takes unity when $a=0$ and $1/2$ when $a=1/2$ (Fig.~\ref{fig:periodicfunction}).

\section{Length and chemical potential asymmetry}
\label{sec:appendix_aymmetric}

When two regions separated by contact points to the two superconductors have different length and chemical potential [Fig.~\ref{fig:analytical} (b)], the eigenenergies are determined by scattering theory of plane waves at two boundaries. 
Let the coordinate of the right contact be denoted by $j=0$ and that of the left contact by $j=N_1$.
Consider plane waves in the two regions
\begin{align}
 \psi(j)
 =
 \left\{
 \begin{array}{ll}
 Ae^{ik_1 j},\quad &(j\in[0,N_1])\\
 Be^{ik_2 j},\quad &(j\in[N_1,N])
 \end{array}
 \right.,
\end{align}
whose Hamiltonian is defined by
$
 \mathcal{H}_{1(2)}
 =
 vk_{1(2)}-\mu_{1(2)}
$.
The momenta for an eigenenergy $\epsilon$ satisfy $vk_1-\mu_1=vk_2-\mu_2=\epsilon$.
Combined with the boundary conditions of the wave function at $j=0$ and $j=N_1$ given by 
$
 A=Be^{ik_2N},Ae^{ik_1N_1}=Be^{ik_2N_1}
$,
the eigenenergies are the same as those of the symmetric system
\begin{align}
 \epsilon_s=\frac{2\pi sv}{N}-\tilde{\mu} \quad (s\in\mathbb{Z}),
\end{align}
where the chemical potential is replaced by $\tilde{\mu}=(\mu_1N_1+\mu_2N_2)/N$.
The corresponding momenta are $k_{1s}=2\pi s/N+(\mu_1-\mu_2)N_2/Nv$ and $k_{2s}=2\pi s/N+(\mu_2-\mu_1)N_1/Nv$.

In the absence of the tunneling to superconductors, the unperturbed Green's function of the electron between the two contact points is given by
\begin{align}
 &[g^\text{edge}_{N_10}(i\omega_n)]_{11}
 =
 \sum_{ss'}
 \psi_s(N_1)\psi_{s'}^{\ast}(0)
 \frac{\delta_{ss'}}{i\omega_n-\epsilon_s} \notag\\
 &=
 \frac{1}{N}
 \sum_{s}
 \exp\left[
 i\frac{\mu_1-\mu_2}{v}\frac{N_1N_2}{N}
 \right]
 \frac{e^{2\pi isN_1/N}}{i\omega_n-\epsilon_s} \notag\\
 &=
 \frac{1}{2v}
 e^{i(\mu_1-\mu_2)N_1N_2/Nv}
 \frac{e^{i(i\omega_n+\tilde{\mu})(N_1-N_2)/2v}}{\sin[N(i\omega_n+\tilde{\mu})/2v]}, \\
 &[g^\text{edge}_{0N_1}(i\omega_n)]_{11}
 =
 \frac{1}{2v}
 e^{-i(\mu_1-\mu_2)N_1N_2/Nv}
 \frac{e^{-i(i\omega_n+\tilde{\mu})(N_1-N_2)/2v}}{\sin[N(i\omega_n+\tilde{\mu})/2v]}.
\end{align}
The unperturbed Green's function in an asymmetric geometry is slightly different from that in the symmetric geometry (\ref{eq:greensfunction_barechiraledge}) by a phase factor depending on the difference of the chemical potential $\mu_1-\mu_2$.
Substituting the unperturbed Green's function into (\ref{eq:dysonequation_chiraledgerightsuperconductor2}) and (\ref{eq:dysonequation_chiraledgeleftsuperconductor2}), the resulting expression of the Josephson current is (\ref{eq:josephsoncurrent_thirdorder}) where $\mu$ and $\delta\varphi$ are replaced by $\tilde{\mu}$ and $\delta\tilde{\varphi}$ defined in (\ref{eq:phasedifference2}).

\end{document}